\documentclass[aip,amsmath,amssymb,reprint,author-year]{revtex4-1}

\usepackage{graphicx}
\usepackage{dcolumn}
\usepackage{bm}
\usepackage[utf8]{inputenc}
\usepackage[T1]{fontenc}
\usepackage[english]{babel}
\usepackage{mathptmx}
\usepackage{hyperref}
\usepackage{amsmath}

\begin{document}

\title[Electromotive force in turbulent stochastic reconnection]{Generation and Effects of Electromotive Force in Turbulent Stochastic Reconnection}

\author{Natalia Nowak}
\affiliation{Astronomical Observatory, Jagiellonian University, ul. Orla 171, 30-244 Krakow, Poland}
\author{Grzegorz Kowal}
\email{grzegorz.kowal@usp.br.}
\author{Diego A. Falceta-Gonçalves}
\affiliation{Escola de Artes, Ciências e Humanidades, Universidade de São Paulo, Av. Arlindo Béttio, 1000 -- Vila Guaraciaba, CEP: 03828-000, São Paulo -- SP, Brazil}

\date{\today}

\begin{abstract}
Reconnection is an important process that rules dissipation and diffusion of
magnetic energy in plasmas. It is already clear that its rate is enhanced by
turbulence, and that reconnection itself may increase its stochasticity, but the
main mechanism that connects these two effects is still not completely
understood. The aim of this work is to identify, from the terms of the
electromotive force, the dominant physical process responsible for enhancing the
reconnection rate in turbulent plasmas. We employ full three-dimensional
numerical simulations of turbulence driven by stochastic reconnection and
estimate the production and dissipation of turbulent energy and cross-helicity,
the amount of produced residual helicity, and determine the relation between
these quantities and the reconnection rate. We observe the development of the
electromotive force in the studied models with plasma-$\beta = 0.1 - 2$ and the
Lundquist number $S=10^{-5}-10^{-4}$. The turbulent energy and residual helicity
develop in the large-scale current sheet, with the latter decreasing the effects
of turbulent magnetic diffusion. We demonstrate that the stochastic
reconnection, apart from the turbulence, can produce a significant amount of
cross-helicity as well. The cross-helicity to turbulent energy ratio, however,
has no correlation with the reconnection rate. We show that its sufficiently
large value is not a necessary condition for fast reconnection to occur. The
results suggest that cross-helicity is inherent to turbulent fields, but the
reconnection rate enhancement is possibly caused by the effects of magnetic
turbulent diffusion and controlled by the residual helicity.
\end{abstract}


\maketitle
%

\section{Introduction}
\label{sec:intro}

Magnetic reconnection is a fundamental process that controls magnetic energy
removal and dissipation in astrophysical systems. It is crucial for heating,
acceleration of particles, and mass accretion. At large scales the reconnection
rate provided by the standard Sweet--Parker theory \citep{Parker:1957,
Sweet:1958} is negligible, as estimated by the relation $V_\mathrm{rec, SP}
\approx V_\mathrm{A} S_L^{-1/2} \ll V_\mathrm{A}$, where $S_L = L V_\mathrm{A} /
\eta$ is the Lundquist number, $L$ - a longitudinal scale of the reconnecting
flux tube, $V_A$ - the Alfvén speed, and $\eta$ - magnetic resistivity. Given
the high conductivity of plasmas and scales of astrophysical systems, one finds
very low values of $V_\mathrm{rec, SP}$ even for astrophysical scenarios where
reconnection is known to occur at fast rates \citep[see,
e.g.,][]{ZweibelYamada:2009, Yamada_etal.:2010, GonzalezParker:2016,
Lazarian_etal:2020}.

Conceptually, fast reconnection can only be achieved by a mechanism for which
$V_\mathrm{rec}$ does not depend on $S_L$ or, if so, should depend on it weakly
(e.g. logarithmically). The earliest numerical works on fast reconnection using
particle-in-cell (PIC) simulations considered collisionless instabilities or the
Hall effect in the generalized Ohm's law \cite[see, e.g.,][]{ShayDrake:1998,
Shay_etal:1998, Birn_etal:2001, Cassak_etal:2005}. More recently,
\cite{Liu_etal:2017} proposed a modified Sweet--Parker theory demonstrating,
that by controlling the opening angle of the outflow region, the reconnection
rate can be fast. They suggested that the mechanism  responsible for the opening
of the outflow region is tearing/plasmoid instability. The Sweet--Parker model
outflow scale limitation imposed by microphysics has been challenged by
\cite{LazarianVishniac:1999}, for which the outflow width scale is determined by
the scales of turbulence rather than plasma microphysics. This fast reconnection
model in the presence of turbulence was successfully tested in a number of
numerical studies \cite[e.g., ][]{Kowal_etal:2009, Kowal_etal:2012,
Higashimori_etal:2013, Takamoto_etal:2015, Jabbari_etal:2016}.

Due to the ubiquitous nature of turbulence in astrophysical sites, one may
understand fast reconnection as a general process as well. Typically turbulence
is understood as being driven externally, at large scales, such as found in
quiescent giant molecular clouds \cite[see][]{Armstrong_etal:1995,
Padoan_etal:2009, ChepurnovLazarian:2010, FalcetaGoncalves_etal:2015}, however,
it may arise naturally, e.g., by collisionless kink instability
\citep{Markidis_etal:2014}, the shock-induced reconnection
\citep{Bessho_etal:2020}, the Kelvin-Helmholtz vortex-induced reconnection
\citep{Nakamura_etal:2020}, or due to plasmoid and/or Kelvin-Helmholtz
instabilities within the stochastic reconnection \citep{Kowal_etal:2020}.

The interlink between turbulence and fast reconnection is, however, still an
open issue. It requires more studies to understand what makes turbulence an
important mechanism for accelerating reconnection rates. As pointed by
\cite{LazarianVishniac:1999}, one of the possible roles of turbulence would be
the enlargement of the outflow scales of the reconnection region. From a
different perspective, \cite{YokoiHoshino:2011} proposed the cross-helicity,
generated by the turbulent motions, as the key ingredient for fast reconnection,
and that sites of non-vanishing cross-helicity should be those where this
process occurs. Instead of the outflow scale, this model focuses on the
enlargement of the regions of inter-crossing field lines. \cite{Yokoi_etal:2013}
studied the space-temporal distribution of transport coefficients due to
turbulence by means of the mean-field approach. In such idealized model, the
reconnection properties were obtained from the balance between the transport
enhancement due to the turbulent energy and its suppression due to the
cross-helicity. According to \cite{Yokoi_etal:2013}, a possible explanation
would be that the increase of cross-helicity suppresses the growth of turbulent
transport with consequent confinement of magnetic diffusion in smaller regions,
and therefore increasing the magnetic reconnection rate. Under this model, a
self-consistent determination of timescales has also been studied
\citep{Widmer_etal:2019}. Indeed, the importance of cross-helicity for the
turbulent diffusion of magnetic fields has been extensively studied, mostly
focused on the evolution of the dynamo effect. Although cross-helicity is
globally conserved in ideal incompressible fluids \citep{Woltjer:1958}, this is
not the case in turbulent compressible plasmas
\citep[e.g.][]{MarschMangeney:1987, SurBrandenburg:2009, Webb_etal:2014a,
Webb_etal:2014b}.

Recent cross-helicity studies in the solar wind using Parker Solar Probe (PSP)
measurements of the proton density and velocity, together with the measurements
of magnetic field, have shown that the alignment between velocity and magnetic
field at convective scales presents a large number of negative value events,
which is due to the presence of magnetic switchbacks, or brief periods where the
magnetic field polarity reverses \citep{McManus_etal:2020}. Their abundance and
short timescales as seen by PSP in its first encounter is a new front of study,
and their precise origin is still unknown. They argue that these must be local
kinks in the magnetic field and not due to small regions of opposite polarity on
the surface of the Sun, which could be possibly related to stochastic
reconnection, i.e., the reconnection in the presence of weakly stochastic
fluctuations of velocity or magnetic field \cite[as
in][]{LazarianVishniac:1999}.

The intrinsic relation between turbulence and fast reconnection has become
clearer in the past few years by means of several theoretical and numerical
studies \cite[see, e.g.,][]{Eyink_etal:2013, Eyink:2015, JafariVishniac:2019,
Lazarian_etal:2020}. Still, not all aspects of this close relation are
understood thoroughly. Here, we have addressed one of the gaps by studying, from
three-dimensional high resolution magnetohydrodynamic (MHD) numerical
simulations of stochastic magnetic reconnection, the loci and effects of terms
of electromotive force, the turbulent energy, and the cross-helicity for the
reconnection rates. The scope of this work is to determine whether there is a
correlation between the growth of reconnection rates and the local enhancement
of turbulent energy, cross-helicity, or residual helicity.

The manuscript is organized as follows. The theoretical description of
electromotive force is provided in Section~\ref{sec:emf}. In
Sections~\ref{sec:modeling} and \ref{sec:averaging} we describe the numerical
simulations used for this work and the procedure of the field separation into
mean and fluctuating components in order to determine the studied quantities,
respectively. These sections are followed by the description of the main results
in Section~\ref{sec:results}. Finally, in Section~\ref{sec:discussion}, we
present our discussion and main conclusions.


\section{Electromotive force in the context of mean-field theory}
\label{sec:emf}

The electromotive force in the mean-field dynamo has been studied analytically
and numerically for a few decades already \cite[see, e.g.,][]{Moffatt:1978,
KrauseRaedler:1980, Yoshizawa:1990, BrandenburgSubramanian:2005,
SchrijverSiscoe:2009, HughesTobias:2010, Charbonneau:2014, Brandenburg:2018}.
The mean-field approach has been also applied to turbulent magnetic reconnection
\citep{YokoiHoshino:2011, Yokoi_etal:2013, Higashimori_etal:2013,
Widmer_etal:2019}.  Here, we briefly describe parts relevant to this work
following \cite{Yokoi_etal:2013}, where the production, transport and diffusion
of turbulent energy, cross-helicity, and residual helicity and their relation to
the electromotive force has been investigated analytically in details.

We consider a system for which the evolution is governed by non-ideal
compressible isothermal MHD equations:
\begin{equation}
\frac{\partial \rho}{\partial t} + \nabla \cdot \left( \rho \vec{u} \right) = 0,
\label{eq:mhd_1}
\end{equation}
\begin{equation}
\begin{split}
\frac{\partial \left( \rho \vec{u} \right)}{\partial t} + \nabla \cdot \left[ \rho \vec{u} \vec{u} + (a^2 \rho + \frac{B^2}{2 \mu_0}) I - \frac{\vec{B} \vec{B}}{\mu_0} \right] = \\ \nu \rho \left[ \nabla^2 \vec{u} + \frac{1}{3} \nabla \left( \nabla \cdot \vec{u} \right) \right],
\end{split}
\label{eq:mhd_2}
\end{equation}
\begin{equation}
\frac{\partial \vec{B}}{\partial t} - \nabla \times \left( \vec{u} \times \vec{B} - \eta \vec{J} \right) = 0,
\label{eq:mhd_3}
\end{equation}
where $\rho$, $\vec{u}$, and $\vec{B}$ are plasma density, velocity and magnetic
field, respectively, $\vec{J} \equiv \nabla \times \vec{B}$ is the current
density, $I$ is the identity matrix, and $a$, $\mu_0$, $\nu$, and $\eta$
describe the isothermal speed of sound, the magnetic permeability constant, the
viscosity, and the magnetic resistivity, respectively. Using the Reynolds
decomposition applied to all fields,
\begin{equation}
  \rho \rightarrow \bar{\rho} + \rho', \, \vec{u} \rightarrow \vec{U} + \vec{u}', \vec{B} \rightarrow \vec{B} + \vec{b}',
\end{equation}
where $\bar{\rho}$, $\vec{U}$, $\vec{B}$ describe the mean density, velocity,
and magnetic field, respectively, and $\rho'$, $\vec{u}'$, and $\vec{b}'$ their
corresponding fluctuating parts, one can separate the evolution of the mean and
fluctuating fields. In the equations governing the evolution of mean fields,
however, will appear terms resulting from the averaging of the high order
products of fluctuating fields discussed in more details below. For convenience,
Eqs.~\ref{eq:mhd_1}-\ref{eq:mhd_3} and the equations describing the mean and
fluctuating fields evolution can be expressed in dimensionless units once the
density is given in the units of $\rho_0$, the velocity and magnetic field in
the units of Alfvén speed, $V_A = B_0 / \sqrt{\mu_0 \rho_0}$, being $B_0$ the
characteristic magnetic field strength. Under this normalization, time is
expressed in the units of Alfvén time $t_A$.

The main quantities involved in the studies of effects of turbulence on mean
fields are represented by the Reynolds stress tensor ${\cal R} = \langle
\vec{u}' \vec{u}' - \vec{b}' \vec{b}' \rangle$ and the turbulent electromotive
force $\vec{E}_M \equiv \langle \vec{u}' \times \vec{b}' \rangle$, where
brackets $\langle \rangle$ indicate spatial averaging at a given scale defining
the mean components. Using the fluctuating parts one can define the kinetic and
magnetic turbulent energies, $K_{u'} \equiv \frac{1}{2} \langle \vec{u}'^2
\rangle$ and $K_{b'} \equiv \frac{1}{2} \langle \vec{b}'^2 \rangle$,
respectively, cross-helicity $W \equiv \langle \vec{u}' \cdot \vec{b}' \rangle$,
and residual helicity $H \equiv \langle - \vec{u}' \cdot \vec{\omega}' +
\vec{b}' \cdot \vec{j}' \rangle$, where $\vec{\omega}' = \nabla \times \vec{u}'$
and $\vec{j}' = \nabla \times \vec{b}'$.

According to \cite{Yokoi_etal:2013} model the productions of turbulent energy
$P_K$ and cross-helicity $P_W$ are described by two terms, one involving the
electromotive force $\vec{E}_M$ and another involving the Reynolds stress ${\cal
R}^{ij}$, being explicitly
\begin{equation}
  P_K = - \vec{E}_M \cdot \vec{J} - {\cal R}^{ab} \frac{\partial U^a}{\partial x^b},
  \label{eq:pk}
\end{equation}
and
\begin{equation}
  P_W = - \vec{E}_M \cdot \vec{\Omega} - {\cal R}^{ab} \frac{\partial B^a}{\partial x^b},
  \label{eq:pw}
\end{equation}
where $\vec{J} = \nabla \times \vec{B}$ and $\vec{\Omega} = \nabla \times
\vec{U}$ are the mean current density and vorticity, respectively \cite[see
Eqs.~29a and 30a in][]{Yokoi_etal:2013}. The turbulent energy and cross-helicity
are subject to dissipation due to the presence of explicit viscosity and
resistivity \cite[Eqs.~29b and 30b in][]{Yokoi_etal:2013}, described by
\begin{equation}
  \varepsilon_K = \nu \bigg \langle \frac{\partial u'^a}{\partial x^b} \frac{\partial u'^a}{\partial x^b} \bigg \rangle
  + \eta \bigg \langle \frac{\partial b'^a}{\partial x^b} \frac{\partial b'^a}{\partial x^b} \bigg \rangle
\end{equation}
and
\begin{equation}
  \varepsilon_W = \left( \nu + \eta \right) \bigg \langle \frac{\partial u'^a}{\partial x^b} \frac{\partial b'^a}{\partial x^b} \bigg \rangle.
\end{equation}
It should be noted, that cross-helicity $W$, and its production $P_W$ and
dissipation $\varepsilon_W$ are not sign-definite.

The electromotive force $\vec{E}_M$ is important. It describes the effects of
turbulence on the mean field induction equation
\begin{equation}
  \frac{\partial \vec{B}}{\partial t} = \nabla \times \left( \vec{U} \times
  \vec{B} - \eta \vec{J} + \vec{E}_M \right).
  \label{eq:induction}
\end{equation}
In order to close the system of equations for the large-scale fields, it is
necessary to express the electromotive force in terms of large-scale $\vec{B}$
and $\vec{U}$, and their derivatives \cite[see, e.g.,][]{Moffatt:1978,
Yoshizawa:1990, SchrijverSiscoe:2009, HughesTobias:2010, Yokoi_etal:2013}. Under
the assumption of inhomogeneous MHD turbulence \citep{Yoshizawa:1990,
Yokoi:2013, Yokoi_etal:2013} the electromotive force can be expressed as
\begin{equation}
  \vec{E}_M = \bar{\alpha} \vec{B} - \bar{\beta} \vec{J}
  + \bar{\gamma} \vec{\Omega},
  \label{eq:emf}
\end{equation}
where $\bar{\alpha}$, $\bar{\beta}$, and $\bar{\gamma}$ are transport
coefficients determined by the statistical properties of turbulence
\begin{equation}
  \bar{\alpha} = \tau_\alpha H, \, \bar{\beta} = \tau_\beta K, \, \bar{\gamma}
  = \tau_\gamma W,
  \label{eq:coeffs}
\end{equation}
where $\tau_\alpha$, $\tau_\beta$, and $\tau_\gamma$ are turbulent time scales
of residual helicity $H$, turbulent energy $K = K_{u'} + K_{b'}$, and
cross-helicity $W$, respectively.  Inserting Eq.~(\ref{eq:emf}) into
Eq.~(\ref{eq:induction}) gives
\begin{equation}
  \frac{\partial \vec{B}}{\partial t} = \nabla \times \left[ \vec{U} \times
  \vec{B} - \left( \eta + \bar{\beta} \right) \vec{J} \right] + \nabla \times
  \left( \bar{\alpha} \vec{B} + \bar{\gamma} \vec{\Omega} \right).
  \label{eq:induction_new}
\end{equation}
The last term represents the generation mechanism of the mean field $\vec{B}$,
while $\bar{\beta}$ acts as the localized turbulent magnetic diffusivity effect.

The dynamical balance between the turbulent energy and cross-helicity is
expected to result in fast reconnection \citep{YokoiHoshino:2011,
Yokoi_etal:2013}.  \cite{YokoiHoshino:2011} demonstrated analytically that the
reconnection rate can be significantly enhanced if the ratio of the
cross-helicity to the total turbulent energy $|W|/K$ is larger than a threshold
of 0.1. Furthermore, \cite{Yokoi:2018a,Yokoi:2018b} considered the effects of
compressibility on the electromotive force which additionally enhances the
reconnection rate \cite[see also][]{Widmer_etal:2019}. The mean-field model by
\cite{Yokoi_etal:2013} has been confirmed to be valid by comparing it to high
resolution simulations of plasmoid instability with a filtering procedure
applied \citep{Widmer_etal:2016b}.


\section{Numerical modeling}
\label{sec:modeling}

The numerical experiments planned in this work were performed in order to
identify the generation and evolution of the terms mentioned above in a
reconnection generated current sheath. Differently to previous works though, we
have studied the problem of electromotive force generation and evolution from
the full numerical solution of the non-ideal, compressible, 3-dimensional MHD
equations (Eqs.~\ref{eq:mhd_1}--\ref{eq:mhd_3}). The domain is defined as a 3D
rectangular domain with physical dimensions $L \times 4L \times L$ (assuming $L
= 1$), the base resolution of $32 \times 128 \times 32$, and the local mesh
refinement up to $5$ levels resulting in the effective resolution of $512 \times
2048 \times 512$ (the effective grid size being $h = 1 / 512$ along all
directions). The refinement criterion was based on the normalized value of
vorticity and current density to follow the turbulence development. We set
periodic boundaries along the X and Z directions, and non-reflecting open
boundaries along the Y direction, at which we apply the condition of the
derivative normal to the boundary plane equal to zero. By placing the Y
boundaries far from the initial current sheet, we guarantee that the boundary
conditions do not affect the development of turbulence, or the reconnection
process itself.

As initial conditions, for the reconnection to be ignited, we consider a Harris
current sheet configuration with a guide field, i.e. $\vec{B} = \left[ \tanh
\left( y / \delta \right), 0, B_g \right]$, as in our previous works
\citep{Kowal_etal:2017, Kowal_etal:2020}.  The density profile was set to
$\rho(y) = 1 + \frac{1}{\beta \left( 1 + B_g^2 \right)} \mathrm{sech}^2 \left( y
/ \delta \right)$ to make the total pressure uniform initially. The isothermal
speed of sound is given by $a = \sqrt{\frac{\beta}{2} \left( 1 + B_g^2
\right)}$, where $\beta = p/p_{mag}$ is the plasma-$\beta$ parameter, and $p =
a^2 \rho$ and $p_{mag} = B^2/2$ are thermal and magnetic pressures,
respectively.  We used Prandtl number $Pr = \frac{\eta}{\nu} = 1$ with viscosity
and resistivity equal $\eta = \nu = S^{-1}$ (see Eqs.~\ref{eq:mhd_2} and
\ref{eq:mhd_3} for explicit viscous and resistive terms). The initial thickness
of the current sheet was set to $\delta = 3.16 \times 10^{-3}$ and the uniform
guide field to $B_g = 0.1$.  The velocity field was initiated with fluctuations
represented by $100$ Fourier modes of random phases and directions, the
amplitude of each mode equal to $10^{-3}$ and wavenumber $k \approx 64 \pi$. The
random velocity fluctuations were set in a narrow region within the vertical
distance up to 0.09 from the initial current sheet. This was done by applying a
window function to the perturbations generated from the Fourier modes, defined
by a function along the $y$ coordinate only, $f(y) = \cos^2{\left(\frac{\pi}{2}
\zeta \right)}$, where $\zeta = \mathrm{min}\left( 1, \mathrm{max} \left( 0,
(y-0.05)/0.04 \right) \right)$. The window function produces a step function
with the transition of a given width smoothed by squared cosine. Therefore, the
initial setup is controlled by the upstream plasma-$\beta$, the current sheet
thickness $\delta$, the guide field strength $B_g$, and the Lundquist number
$S$.

The MHD equations (\ref{eq:mhd_1})--(\ref{eq:mhd_3}) were solved using a
high-order shock-capturing adaptive refinement Godunov-type code {\tt
AMUN} (\href{https://bitbucket.org/amunteam/amun-code}{https://bitbucket.org/amunteam/amun-code})
with 5$^{th}$-order Optimized Compact Monotonicity
Preserving reconstruction of Riemann states \citep{AhnLee:2020}, the HLLD
Riemann flux solver \citep{Mignone:2007}, and the 3rd-order 4-step Strong
Stability Preserving Runge-Kutta (SSPRK) method \citep{Gottlieb_etal:2011} for
time integration. Due to the initial upstream density and the strength of
reconnecting component of magnetic field being unities they resulted in the
velocity and simulation time units $[ v ] = V_A = 1$ and $[t] = t_A = L / V_A =
1$, respectively.

We present here results for three models with different values of plasma-$\beta$
and Lundquist number $S$: $\beta = 0.1$ and $S=10^5$ (model A), $\beta = 2$ and
$S=10^5$ (model B) and $\beta = 2$ and $S=10^4$ (model C) run for $10-30$ Alfvén
times.

\begin{figure*}[t]
 \centering
 \includegraphics[width=0.49\textwidth]{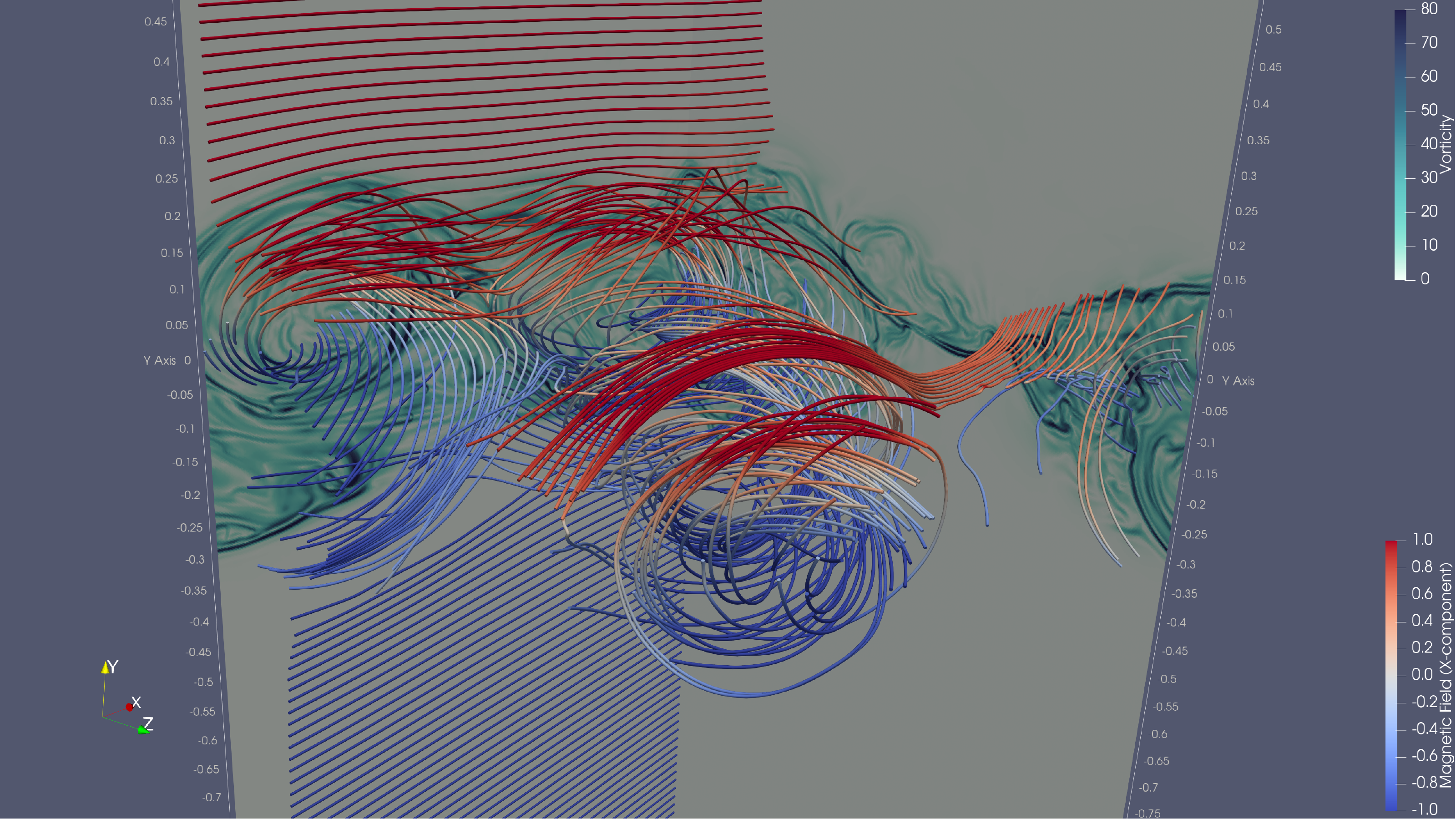}
 \includegraphics[width=0.49\textwidth]{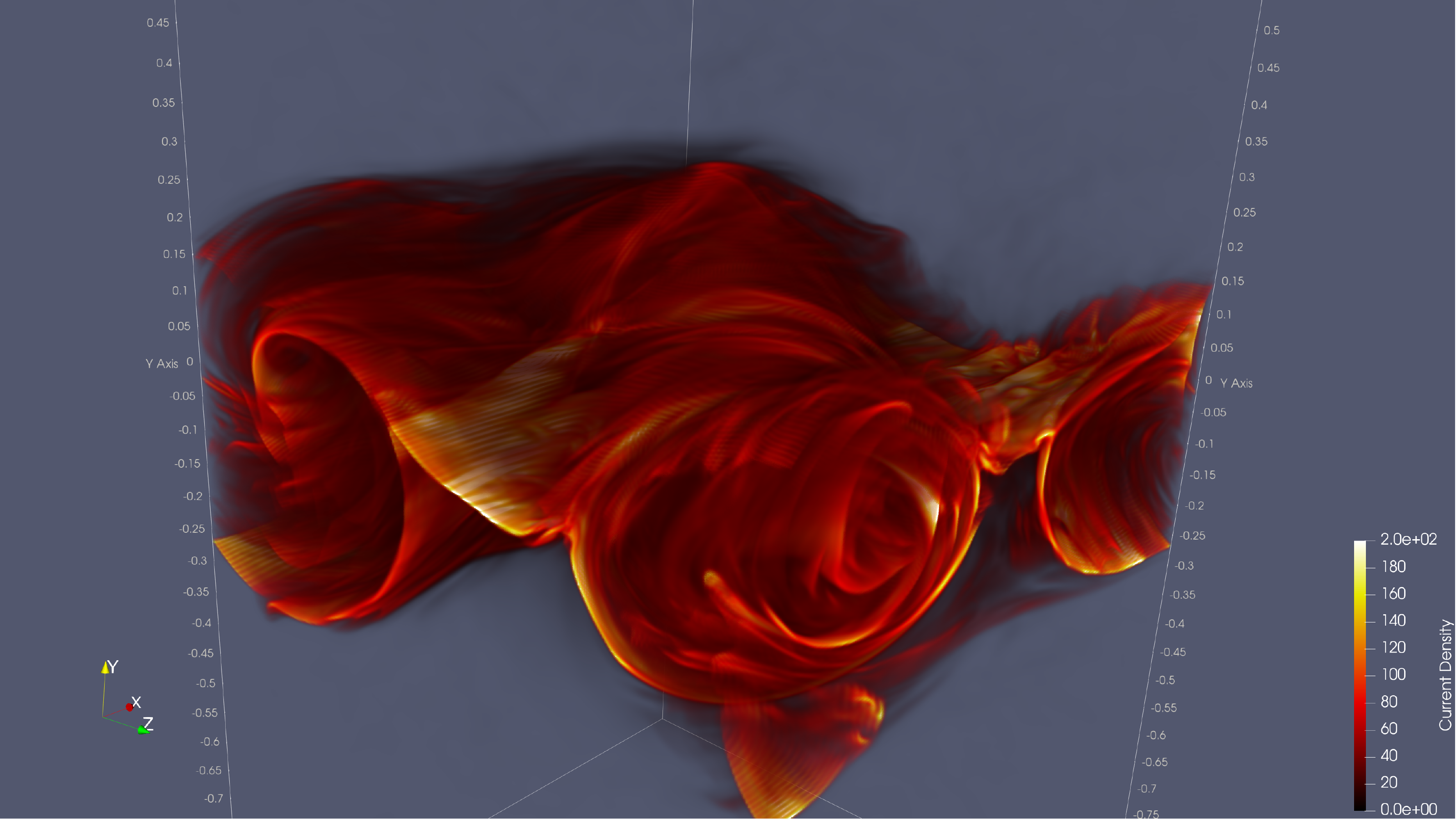}
 \caption{Visualization of the selected magnetic lines and vorticity projected
on the computational domain boundaries (left plot) and the current density
intensity (right plot) for model A at the final moment $t=10$. Only the region
of the computational domain where the turbulence is developed is shown, from
around -0.5 to 0.5 along the Y-direction. The full domain extends from -2 to 2
along this direction.}
 \label{fig:visualization}
\end{figure*}


\section{Field Separation Procedure}
\label{sec:averaging}

Once the numerical simulations are run, for the several data sets of different
timescales, we want to determine the electromotive force, turbulent energy and
cross-helicity terms. In order to do so, as explained in \S\ref{sec:emf}, we
must decompose the mean and fluctuating terms of the variables evolved.

The separation which guarantees the self-consistent determination of both mean
and fluctuating fields is the Reynolds decomposition. It satisfies a certain set
of rules known as Reynolds rules, i.e. $\langle c f \rangle = c \langle f
\rangle$, $\langle f + g \rangle = \langle f \rangle + \langle g \rangle$,
$\langle f' \rangle = 0$, $\langle \langle f \rangle \rangle = \langle f
\rangle$, $\langle \langle f \rangle \cdot g \rangle = \langle f \rangle \langle
g \rangle$, $\langle f' \cdot g' \rangle = \langle f \cdot g \rangle - \langle f
\rangle \cdot \langle g \rangle$, etc., where $\langle \rangle$ means the
ensemble average, $f,g$ are arbitrary fields, $f' = f - \langle f \rangle$, $g'
= g - \langle g \rangle$ are the fluctuating parts, and $c$ is a constant. The
equations presented in \S\ref{sec:emf} were derived under the assumption of the
Reynolds decomposition.

In this work, the mean fields were determined by spatial averaging over the
$XZ$-planes. This type of averaging satisfies the Reynolds rules and is adequate
to the symmetry of our problem, which is characterized by an initial sharp
transition of the magnetic field across the $y=0$ plane and near uniformity far
from the turbulent region during the whole evolution. Since our simulations are
based on the adaptive mesh, before the averaging was applied, we used trilinear
interpolation to represent the fields on the uniform mesh equivalent to the
effective resolution. By subtracting the large-scale components from the
original fields we obtained the fluctuating parts $\vec{u}'(x, y, z)$ and
$\vec{b}'\left( x, y, z \right)$. The averages of products of fluctuating parts
were determined using the corresponding rule listed in the previous paragraph.
They determine the electromotive force $\vec{E}_M$ and other related quantities.
The described procedure produces mean fields varying solely along the
Y-direction. This results in the Y-components of the mean vorticity and current
density to be zero. Additionally, in order to fulfill the divergence-free
condition of the mean $\vec{B}$ we assumed its $Y$-component to be zero. It
should be clarified, that the averaging procedure applied to the initial
velocity fluctuations results in the mean vorticity $\vec{\Omega}$ which is not
zero at $t=0$. $\vec{\Omega}$ reaches very small amplitudes, slightly above
$5\times10^{-3}$ near the initial current sheet in our setup.

The applied averaging procedure dictates the scope of our approach, which is to
study the response of the global magnetic reconnection process, defined by the
profiles along the $y$ coordinate, to the presence of longitudinal (with respect
to the current sheet plane) turbulent fluctuations produced by the reconnection
itself. Clearly, the short-wave fluctuations of the mean fields along the
Y-direction are not perfectly averaged by this procedure, as explained above.
Nevertheless, by using the periodicity along the X and Z directions, we
guarantee that the variations of the mean fields along the Y-direction is the
result of the initial velocity perturbation and developed turbulence.

We should remark that the averaging over the XZ-planes can affect the
interpretation of our results in the context of \cite{Yokoi_etal:2013} model,
which was derived under the assumption of inhomogeneous turbulence. Still, the
analytical framework of their model was an inspiration for the studies presented
in this manuscript. Our averaging procedure treats any kind of variability along
the X or Z directions as the fluctuating part. We do not analyze the horizontal
profiles of the quantities which are not sign-definite, such as cross-helicity.
We plan to apply to our simulations alternate averaging procedures, such as
Gaussian filtering \cite[see, e.g., \S2 in][]{Sagaut:2006}, in the future.


\section{Results}
\label{sec:results}

In this work we study the evolution of electromotive force $\vec{E}_M$, the
turbulent energy $K$, its production $P_K$ and dissipation $\varepsilon_K$, the
cross-helicity $W$, its production $P_W$ and dissipation $\varepsilon_W$, the
residual helicity $H$, the scalar product $\vec{J}\cdot \vec{B}$, and the ratio
$|W|/K$, using the mean-field approach inspired by the model presented in
\cite{Yokoi_etal:2013}, in stochastic turbulent reconnection
\citep[see][]{Kowal_etal:2017, Kowal_etal:2020}. Moreover, we investigate the
causality between these quantities and the reconnection rate $V_{rec}$ measured
in these models.

In Figure~\ref{fig:visualization} we show an example visualization of the model
A at the final time $t = 10$. In the left plot, we show the selected magnetic
lines with the red and blue colors identifying their orientation with respect to
the X-axis (red for the parallel and blue indicating the anti parallel
orientations with respect to the X axis). As seen by the degree of the line
wandering, the magnetic field above and below the turbulent region is relatively
laminar. The oppositely oriented field lines are mixed in the turbulent region
resulting in a complex current density structure, seen in the right plot of
Figure~\ref{fig:visualization}. Together with the magnetic lines we show two
cuts of the vorticity distribution in the left plot, one along the XY and
another along the YZ plane, which identify the turbulent region. Clearly, the
developed turbulence directly affects the reconnection process, as well as, its
rate.

\begin{figure*}[t]
 \centering
 \includegraphics[width=0.32\textwidth]{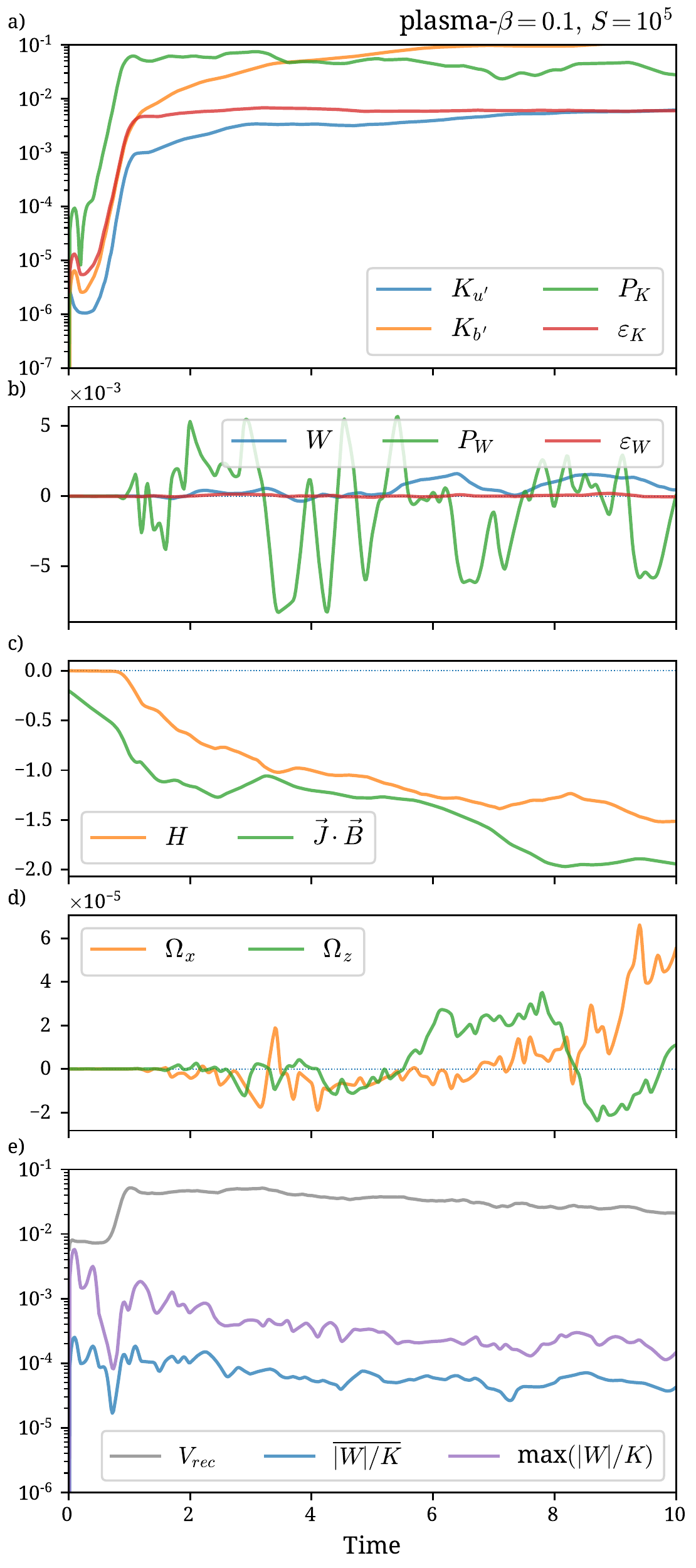}
 \includegraphics[width=0.32\textwidth]{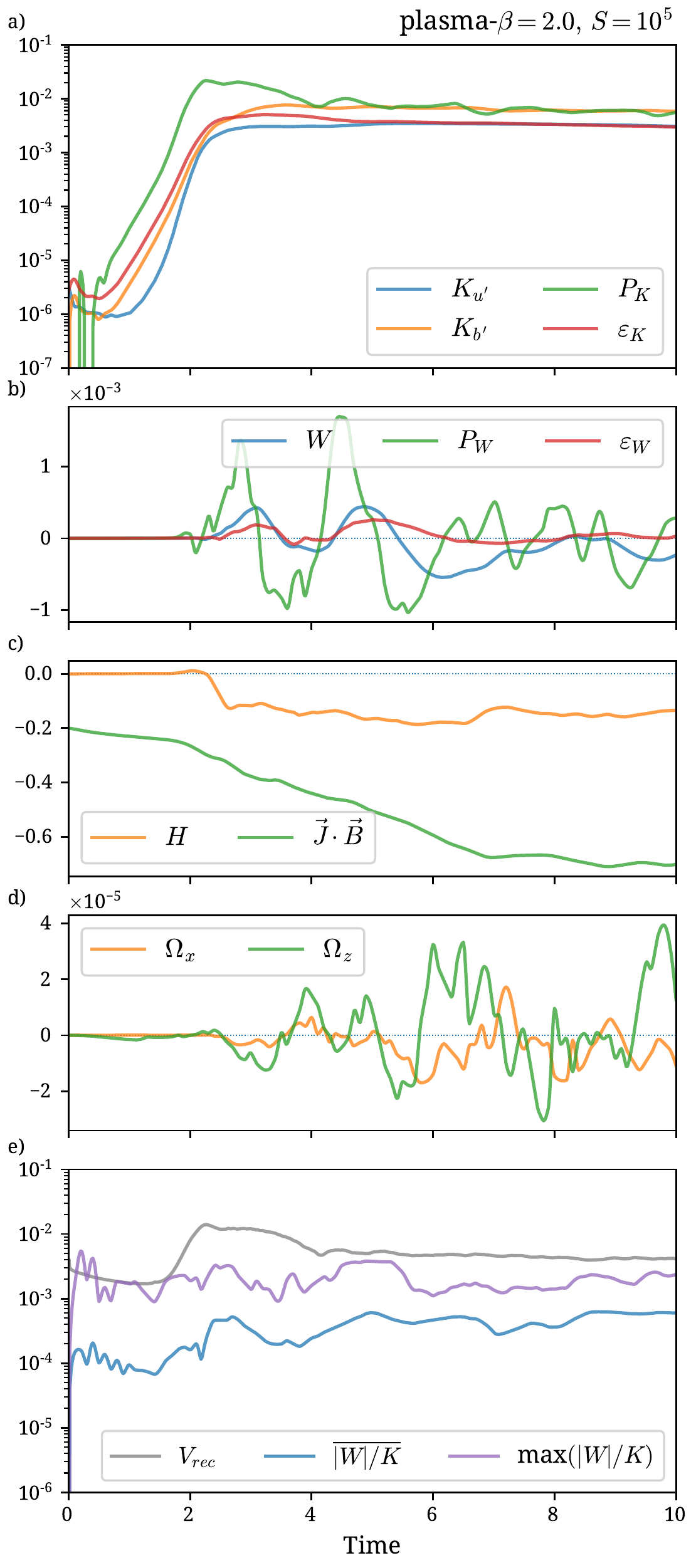}
 \includegraphics[width=0.32\textwidth]{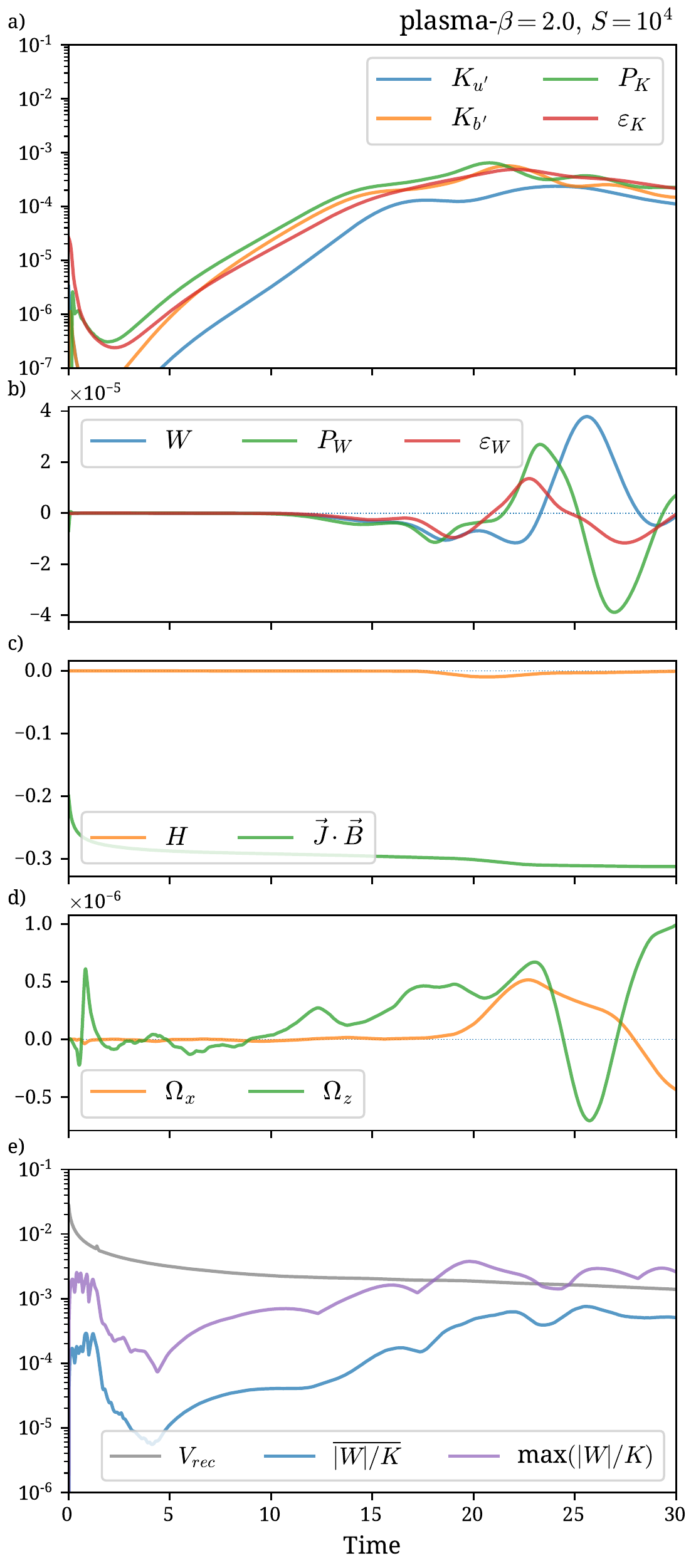}
 \caption{Evolution of (a) the kinetic $K_{u'}$ and magnetic $K_{b'}$ turbulent
energies, the production $P_K$ and dissipation $\varepsilon_K$ of turbulent
energy, (b) the cross-helicity $W$, its production $P_W$ and dissipation
$\varepsilon_W$, (c) the residual helicity $H$ and the scalar product
$\vec{J} \cdot \vec{B}$, (d) the large-scale vorticity components,
$\Omega_x$ and $\Omega_z$, and (e) the mean ratio of cross-helicity to total
turbulent energy $\overline{|W|/K}$ and the reconnection rate $V_{rec}$, for
models A, B, and C ({\em left}, {\em middle}, and {\em right}, respectively).
All quantities shown in plots a)-d) are integrated over the computational
domain. The mean value of the ratio $|W|/K$ was calculated over a region near
the mid-plane containing 99\% of total turbulent energy (see
Fig.~\ref{fig:profiles}).}
\label{fig:evolution}
\end{figure*}

In Figure~\ref{fig:evolution} we present time evolution of turbulent energies
$K_{u'}$ and $K_{b'}$, the turbulent energy production $P_K$ and dissipation
$\varepsilon_K$ ({\em a}), the cross-helicity $W$, its production $P_W$ and
dissipation $\varepsilon_W$ ({\em b}), the residual helicity $H$ and the scalar
product $\vec{J} \cdot \vec{B}$ ({\em c}), the large-scale components of
vorticity $\Omega_x$ and $\Omega_z$ ({\em d}), all integrated over the $y$
coordinate. In addition, in the last row (e) of Figure~\ref{fig:evolution}, we
show the maximum value of the ratio of the absolute value of cross-helicity over
total turbulent energy, $|W|/K$, as well as, the mean value averaged over the
mid-plane region containing 99\% of the total turbulent energy and the
reconnection rate $V_{rec}$ determined using the method described in
\cite{Kowal_etal:2009}. The estimation of $V_{rec}$ is based on the conservation
of the magnetic flux related to the reconnecting component $B_x$. It tracks the
temporal variation of the absolute value of $B_x$ integrated over the
computational domain taking into account the field brought to the system through
the open Y-boundaries, $V_{rec} = \left[ \iint_A{U_y |B_x| \, dxdz} -
\frac{d}{dt} \iiint_V{|B_x| \, dxdydz} \right] / B_0$, where $A$ is the Y
boundary area, $V$ is the computational domain volume, and $B_0$ is the
amplitude of $B_x$ averaged over $A$. Since our Y-boundaries are far from the
current sheet and the region of developed turbulence, the terms related to the
shear and dissipation of magnetic field at the boundaries could be neglected
\cite[see \S4 in][for more details]{Kowal_etal:2009}.

In all models shown in Figure~\ref{fig:evolution} the evolution can be divided
into two stages: the growth and the saturation phases. Focusing on the growth
phase first, we see that its duration depends both on plasma-$\beta$ and the
Lundquist number $S$. Comparing the evolution of kinetic turbulent energy
$K_{u'}$, the estimated growth rates (within the period of exponential growth)
are $~12.4$, $~6.2$, and $~0.7$ for model A, B, and C, respectively, indicating
that both lower plasma-$\beta$ and higher Lundquist number favor the quicker
development of turbulence in stochastic reconnection. The magnetic turbulent
energy is characterized by similar growth rates: $~12.1$, $~5.2$, and $~0.7$,
for model A, B, and C, respectively. This suggests a close interaction of
velocity and magnetic eddies exchanging the turbulent energy between them. Apart
from the increase of turbulent energies, we also observe a similar growth of the
production and dissipation of turbulent energy (Fig.~\ref{fig:evolution}a, green
and red lines, respectively).

It is interesting to see that, during the growth phase, there is essentially no
generation of cross-helicity $W$ and the residual helicity $H$
(Fig.~\ref{fig:evolution}b, c). The production of cross-helicity $P_W$ and it
dissipation $\varepsilon_W$ are relatively negligible during this stage too.
This indicates that these quantities start to be generated once the full
turbulent cascade is formed. As shown in \citep{Kowal_etal:2017}, the cascade is
build up from the small scales toward the large scales during the growth phase
(inverse cascade).

The only variable sensitive to the growth of the turbulent energies, except the
turbulent energy, is the reconnection rate $V_{rec}$, however, only in the high
Lundquist number models A and B. There, the growth of the reconnection rate is
visible exactly during the growth phase observed in the turbulent energy
evolution. In model C, even though the turbulent energy increases by over two
orders of magnitude, $V_{rec}$ constantly decays.

The fact that the reconnection rate reaches larger values in the low
plasma-$\beta$ model, and the turbulence develops faster in this regime, could
be explained by higher degree of compressibility. Low plasma-$\beta$ means that
plasma motions can be mildly or even weakly supersonic in the vicinity of the
current sheet, increasing the compression of plasma, which in turn reduces the
importance of magnetic tension. This results in easier mixing of the magnetic
flux within the turbulent region, and more efficient turbulent dissipation
($\bar{\beta}$-effect). If the explicit diffusion (viscosity and resistivity) is
sufficiently large, however, it dissipates the fluctuations at scales comparable
to the current sheet thickness on time scales shorter than the reconnection time
scale, and cannot sufficiently affect the reconnection rate. However, the
fluctuations at scales $l > \delta$ can affect the shape of the current sheet,
which ejects the kinetic energy from the reconnection outflow regions.

Analyzing the evolution of turbulent energies, $K_{u'}$ and $K_{b'}$ (blue and
orange, respectively) during the saturation phase in
Figure~\ref{fig:evolution}a, we see that in the case of high Lundquist number
models A and B (left and middle), the kinetic turbulent energy $K_{u'}$ reaches
values about $5.9 \times 10^{-3}$ and $3.2 \times 10^{-3}$, respectively, while
in the case of model C (right) this energy is over an order of magnitude
smaller, slightly above $10^{-4}$ at the end of the simulation. The final
turbulent magnetic energy $K_{b'}$ level strongly depends on the values of
plasma-$\beta$ and $S$, reaching over $10^{-1}$, $\sim 6 \times 10^{-3}$, and
$\sim 1.5 \times 10^{-4}$ for models A, B, and C, respectively. In model A the
turbulent magnetic energy is nearly an order of magnitude larger than the
kinetic one, while in models with plasma-$\beta = 2$ they are nearly in
equipartition with $K_{u'}$ slightly smaller than $K_{b'}$. In the lower
Lundquist number model C the kinetic and magnetic energies reach the maximum
level of over $10^{-4}$ around time $t=15$, however, they start to slowly
decrease afterwords. The production and dissipation of turbulent energy are
shown in Figure~\ref{fig:evolution}a with green and red lines, respectively.
$P_K$ saturates at levels around $4 \times 10^{-2}$ and $5.6 \times 10^{-3}$, a
difference of around 7 times, for models A and B, respectively. It is
interesting that the dissipation $\varepsilon_K$ for these models reaches
somewhat less discrepant levels, of around $6 \times 10^{-3}$ and $3 \times
10^{-3}$, respectively. Significantly different levels of $P_K$ and
$\varepsilon_K$ suggest continuous production of turbulent energy, as can be
seen clearly in the case of low plasma-$\beta$ model A, where the levels of
$K_{u'}$ and $K_{b'}$ increase gradually even after reaching the saturation. In
model C, the production $P_K$ and dissipation $\varepsilon_K$ increase until
time about $t=20$, with $P_K$ being larger than $\varepsilon_K$ by a factor of
around 2. After $t=20$, $P_K$ and $\varepsilon_K$ start to decay with the
dissipation slightly overcoming the production. In all these cases, the positive
balance of $P_K$ over $\varepsilon_K$ indicates the growth of turbulent energy.
Normally, the turbulent energy or cross-helicity could be produced in one place
and then transported through out the domain. Due to the fact that we assume
large-scae $B_y = 0$ to guarantee the divergence-free large-scale magnetic
field, the large-scale transport, expressed by the $\vec{B} \cdot \nabla$ terms
\citep[Eqs.~29c and 30c in][]{Yokoi_etal:2013}, is therefore null. This seems to
be consistent with the visualizations presented in
Figure~\ref{fig:visualization}, which demonstrate that indeed the turbulent
region is constrained in the vertical direction. Therefore, the net production
of the turbulent energy and cross-helicity are determined by $P_K -
\varepsilon_K$ and $P_W - \varepsilon_W$, respectively.

In the second row of Figure~\ref{fig:evolution} we show the evolution of the
cross-helicity $W$ (blue), its production $P_W$ (green) and dissipation
$\varepsilon_W$ (red) integrated over the computational domain. We see that $W$
oscillates around zero with an increasing amplitude reaching around $10^{-3}$
and $5\times10^{-4}$ at later times for models A and B (left and middle panels),
respectively. The mean values averaged over the last 4 time units for these
models are $7 \times 10^{-4}$ and $-2 \times 10^{4}$, respectively. The periods
of increasing and decreasing cross-helicity are well correlated with the sign of
the production $P_W$ in all models. It is interesting to see that the
dissipation of cross-helicity $\varepsilon_W$ is much weaker in models with
$S=10^5$ indicating that the production $P_W$ is mainly responsible for the
oscillating character of $W$. In the case of model C (right panel) the
cross-helicity tends to negative values, significantly weaker in terms of
amplitudes (by two orders of magnitude) when comparing to models A or B. The
dissipation $\varepsilon_W$ is only weaker by a factor of around 2 comparing to
the production in this case, in contrast to other models, most probably due to
large values of viscosity and resistivity.

\begin{figure*}[t]
 \centering
 \includegraphics[width=0.32\textwidth]{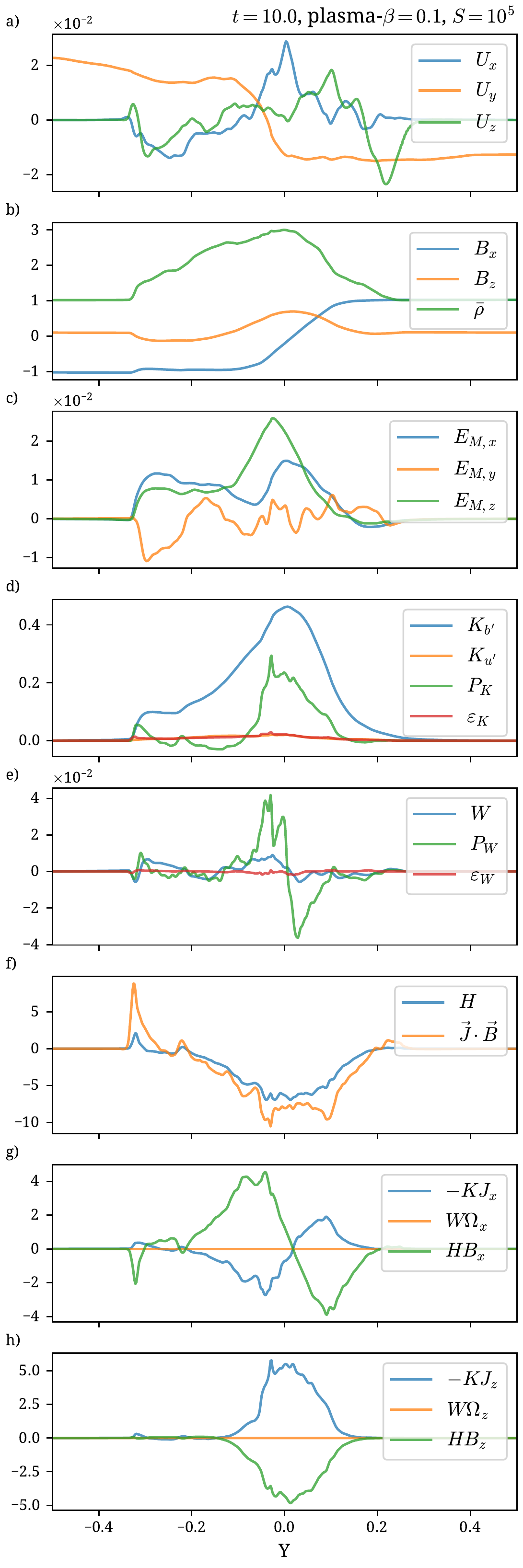}
 \includegraphics[width=0.32\textwidth]{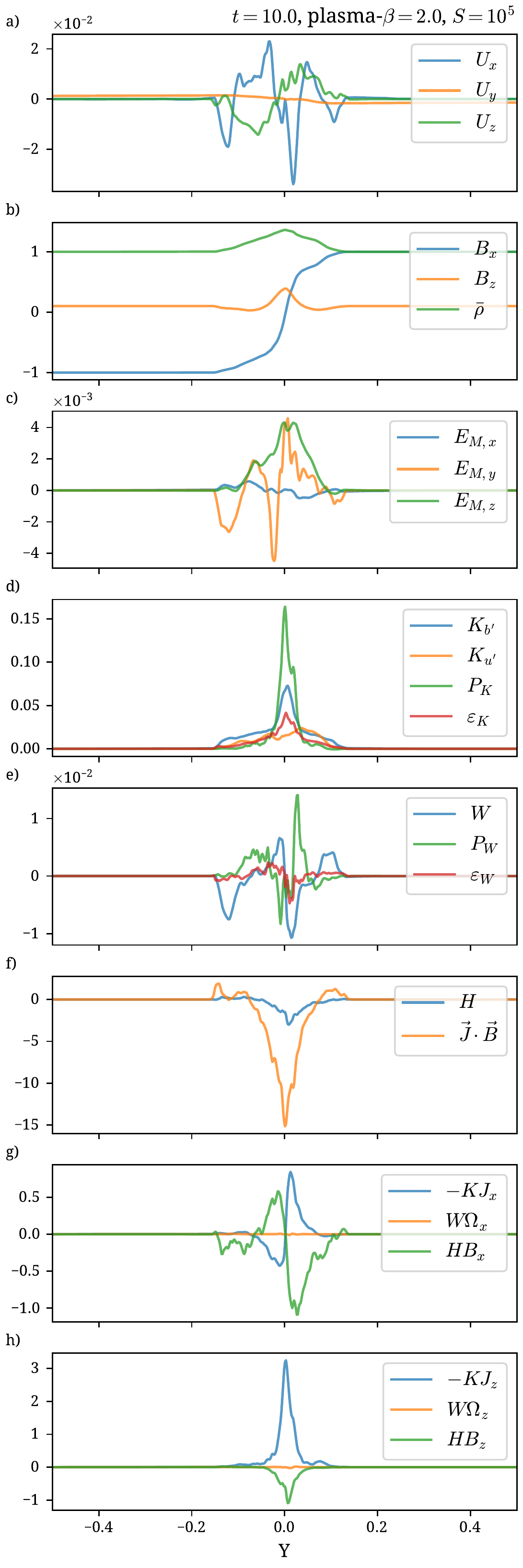}
 \includegraphics[width=0.32\textwidth]{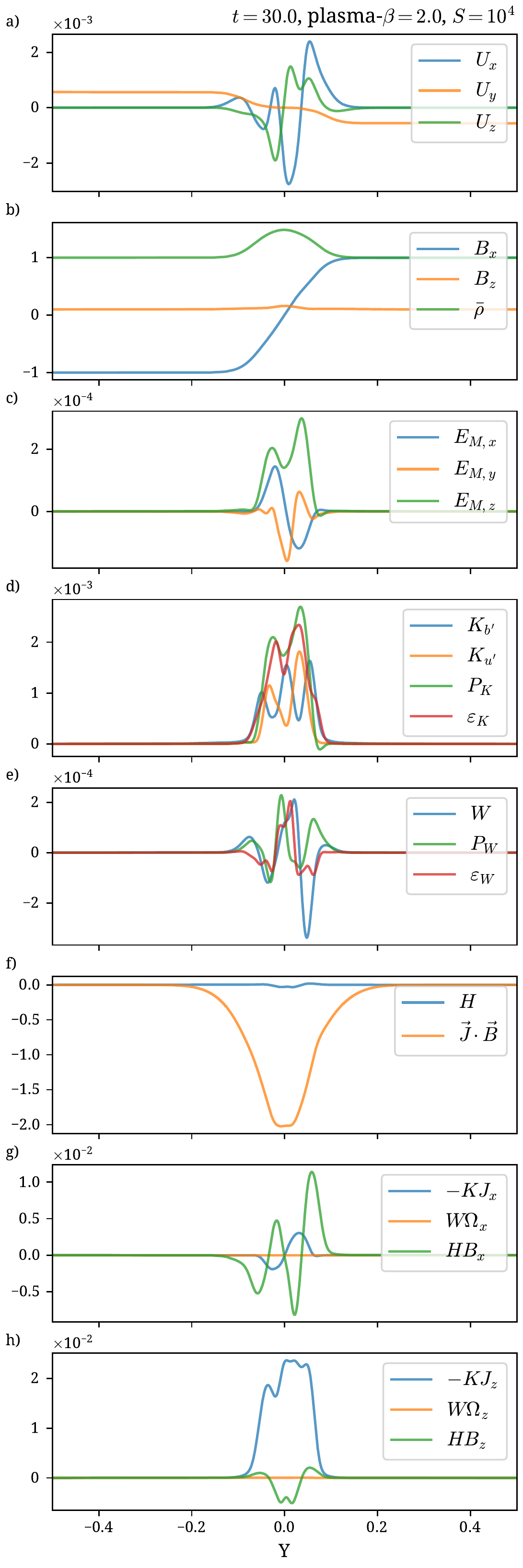}
 \caption{From top to bottom, profiles of (a) the mean velocity $\vec{U}$
components, (b) the mean magnetic field components ($B_x$ and $B_z$) and density
$\bar{\rho}$, (c) the electromotive force $\vec{E}_{M}$ components, (d) the
kinetic $K_{u'}$ and magnetic $K_{b'}$ turbulent energies, the production $P_K$
and dissipation $\varepsilon_K$ of turbulent energy, (e) the cross-helicity $W$
and its production $P_W$ and dissipation $\varepsilon_W$, (f) the residual
helicity $H$ and the scalar product of the large-scale current density and
magnetic field $\vec{J}\cdot\vec{B}$, (g) the comparison of terms $-K J_x$, $W
\Omega_x$, and $H B_x$, and (h) the comparison of terms $-K J_z$, $W \Omega_z$,
and $H B_z$, for models A, B, and C ({\em left}, {\em middle}, and {\em right},
respectively) at the final simulation times.}
\label{fig:profiles}
\end{figure*}

In the third row of Figure~\ref{fig:evolution} we show the evolution of the mean
residual helicity $H$.  This quantity seems to be strongly related to
plasma-$\beta$. It grows gradually to negative values reaching around $-1.5$ at
the final time of simulation in the case of model A. For model B, after a very
short period of small positive values at around $t=2$, it quickly decays to
values below $-0.1$ and oscillates around $-0.15$. The model C manifests much
weaker and more delayed development of $H$. After long initial oscillations
around zero, with a very weak amplitude, it quickly decreases to $-8 \times
10^{-3}$ only after $t = 18$ and returns to zero at the later times. For
comparison, we show in the same plot the evolution of the integrated quantity
$\vec{J} \cdot \vec{B}$, which, according to Eq.~43 in \cite{Widmer_etal:2016a},
determines the evolution of $\bar{\alpha}$. The evolution of $\vec{J} \cdot
\vec{B}$ resembles the evolution of $H$ only in the low plasma-$\beta$ model A.
In model B, the scalar product of large-scale current density and magnetic field
constantly decreases saturating only after about $t=7$ at the value around
$-0.7$, much lower than $H$ which saturates much earlier. In the case of model
C, however, $\vec{J} \cdot \vec{B}$ reaches values around $-0.3$ very quickly,
as compared to nearly negligible values of $H$.

The next row in Figure~\ref{fig:evolution} (row d) shows the evolution of the
large-scale components of vorticity integrated over the domain. As noted
earlier, the initial vorticity is weak but not zero. As seen in these plots, the
large-scale vorticity fluctuates around zero with amplitude increasing during
the whole evolution, with the Z component dominating over the X one. The
increase of $\Omega_z$ is especially well seen in models B and C. It may be
related to the development of elongated velocity shear structures resulting in
Kelvin-Helmholtz instability already reported in \cite{Kowal_etal:2020}.

In the last row of Figure~\ref{fig:evolution} we show the reconnection rate
$V_{rec}$ and the maximum value of the ratio $|W|/K$ and $\overline{|W|/K}$, the
mean value of the ratio averaged over the region of developed turbulent energy.
The averaging was done around the peak value of turbulent energy along the
$Y$-direction (see Fig.~\ref{fig:profiles}d), within the distance containing
99\% of the total turbulent energy $K$.  The value of $\overline{|W|/K} $
saturates at around $5 \times 10^{-5}$ for model A (left) and around $6 \times
10^{-4}$ for model B (middle). For model C (right) $\overline{|W|/K}$ saturates
at late times at values comparable with model B, indicating a weak dependence on
$S$. The evolution of the maximum of $|W|/K$ is relatively consistent with the
mean value $\overline{|W|/K}$ in terms of the temporal variability, being about
an order of magnitude larger. There is relatively no correlation between the
ratio $|W|/K$ and the reconnection rate $V_{rec}$, indicating that the
enhancement of the reconnection rate observed in models A and B (left and
middle) is not due to the generation or amplification of cross-helicity. In
model C (right), $|W|/K$ grows during the whole simulation until its saturates,
while $V_{rec}$ constantly decays. We believe, however, that these results do
not invalidate \cite{YokoiHoshino:2011} hypothesis about a threshold above which
the cross-helicity is expected to significantly affect the reconnection rate,
given that the estimated values of $|W|/K$ in our models are much lower.

In Figure~\ref{fig:profiles} we present vertical profiles (along the $y$
coordinate, the only spatial dependence due to the result of the averaging
procedure applied) of several quantities studied in this work. From top to
bottom we show, (a) the large-scale components of velocity $U_x$, $U_y$, and
$U_z$, (b) the large-scale components of magnetic field, $B_x$ and $B_z$, and
density $\bar{\rho}$, (c) the components of electromotive force $\vec{E}_M$, (d)
the kinetic and magnetic turbulent energies, $K_{u'}$ and $K_{b'}$,
respectively, the production and dissipation of turbulent energy, $P_K$ and
$\varepsilon_K$, respectively, (e) the cross-helicity $W$ and its production and
dissipation, $P_W$ and $\varepsilon_W$, respectively, (f) the residual helicity
$H$ and the scalar product $\vec{J} \cdot \vec{B}$, and (g) the terms $-K J_x$,
$W \Omega_x$, $H B_x$ and (h) $-K J_z$, $W \Omega_z$, and $H B_z$, contributing
to the generation of the electromotive force components $E_{M,x}$ and $E_{M,z}$,
respectively, according to \cite{Yokoi_etal:2013} model. The columns from left
to right correspond to models A, B, and C, respectively, at their final
simulation times.

The profiles of mean velocity show that the amplitude of $U_y$ (orange in
Fig.~\ref{fig:profiles}a) decreases with increasing plasma-$\beta$. This
component is related to the reconnection rate, since it determines the amount of
magnetic field brought to the system to compensate the reconnected flux. In
order to determine $V_{rec}$ one has to track the amount of magnetic flux within
the system, as described earlier in this section. Indeed, measuring the $U_y$
amplitude close to Y-boundaries, we get values of about $1.8 \times 10^{-2}$,
$8.4 \times 10^{-4}$, and $5.6 \times 10^{-4}$ for models A, B, and C,
respectively, comparing to respective $V_{rec} = 2.3 \times 10^{-2}$, $4.2
\times 10^{-3}$, and $1.4 \times 10^{-3}$ at the final time. As we explained
already, these values depend on the degree of compressibility through the
plasma-$\beta$ parameter, and on the Lundquist number, which lower values result
in efficient dumping of the generated turbulent fluctuations at scales
comparable to the current sheet thickness. Other components reach comparable
amplitudes, independent of plasma-$\beta$. In model C amplitudes of X and Z
components are significantly smaller comparing to model B. We see that the
turbulent region is broader for the low plasma-$\beta$ model A. The width of
this region corresponds to increased large-scale density region seen in the next
row (green line). In these plots we see the roughly maintained region of the
polarization change of $B_x$ (blue) and development of the amplified guide field
$B_z$ (orange).

Analyzing the components of electromotive force (third row in
Fig.~\ref{fig:profiles}), we notice that their amplitudes are sensitive to the
plasma-$\beta$ parameter. In model A the X and Z components reach similar
amplitudes of around $10^{-2}$ with the Y component being the weakest, while in
model B $E_{M,x}$ seems to be the weakest one at the final simulation time.
Model C shows a similar behaviour as model B, however, with $E_{M,x}$ being
comparable to $E_{M,y}$ in the amplitude. In both these models the amplitude of
$E_{M,z}$ is the largest among all components. It also expands in a broader
region around the current sheet as compared to other components. The profiles of
turbulent energies (fourth row in Fig.~\ref{fig:profiles}) clearly show the
evolution of turbulent energies near the current sheet where the reconnection
takes place, with magnetic energy (blue) significantly dominating in the low
plasma-$\beta$ model, probably due to strong deformation of the initial current
sheet. In the same row we show the production and dissipation of turbulent
energy, $P_K$ (green) and $\varepsilon_K$ (red), respectively, peaking at the
midplane. In the low plasma-$\beta$ model A, the dissipation significantly
dominates the production at this moment, both in terms of the amplitude and
volume. In model B, however, the production dominates in a narrow region near
$y=0$. In model C, we observe a much smaller dominance of the production over
dissipation in the region of developed turbulence.

In all models the cross-helicity $W$ (see Fig.~\ref{fig:profiles}e) changes its
sign across the reconnection plane. In the low plasma-$\beta$ model A, at the
final time, the cross-helicity $W$ (blue) reaches values of around $10^{-2}$ and
its production rate $P_W$ (green) a few times larger, both oscillating around
zero. Their amplitudes are slightly bigger than those observed in model B. As
for the distribution of dissipation of cross-helicity $\varepsilon_W$, it has
much weaker amplitude comparing to $P_W$ for model A. It becomes stronger in
model B, while in model C it is nearly as strong as the cross-helicity
production. Still, the values of $\varepsilon_W$ are smaller that those of
$P_W$, indicating that the simulated systems are still able to produce
cross-helicity. For completeness, we also show the distribution of the residual
helicity $H$ (the row {\em f} in Fig.~\ref{fig:profiles}) strongly developing in
the region of current sheet and peaking at negative values in all models. The
profile of corresponding scalar product of the mean current density and magnetic
field, $\vec{J} \cdot \vec{B}$, looks quite similar to the profile of $H$ for
model A. However, in model B and C, $\vec{J} \cdot \vec{B}$ is stronger by an
order of magnitude or more. The guide field seems to play an important role in
the development of residual helicity or $\vec{J} \cdot \vec{B}$, as already
reported in \cite{Widmer_etal:2016a}. As we mentioned already, the profiles of
mean $B_z$ seen in Figure~\ref{fig:profiles}a show the guide field to be
significantly enhanced around the mid-plane. This enhanced guide field together
with already strong $J_z$ due to the global profile of the reconnecting
component $B_x$ contribute mostly to the observed profiles of $\vec{J} \cdot
\vec{B}$. On the other hand, $H$ determines the balance between the kinetic and
magnetic helicities, as it is the result of the scalar product between the
fluctuating parts of velocity and magnetic field, and their curls. Clearly,
strong variability of the fluctuating parts will enhance the residual helicity,
what is especially well seen in the case of for model A. The level of
compressibility can affect the production of $H$, as can be seen by comparing
models A and B (Fig.~\ref{fig:evolution}f, left and central panels), or the
enhanced explicit diffusivity can substantially stop its production
(Fig.~\ref{fig:evolution}f, right panel).

Finally, in the last two rows of Figure~\ref{fig:profiles} we show the profiles
of terms $-K J_i$, $W \Omega_i$, and $H B_i$ (blue, orange, and green,
respectively) where $i=x,z$ for the X and Z components, respectively. These
terms determine the degree of individual contributions to the corresponding
component of the electromotive force $\vec{E}_{M,i}$ from turbulent energy,
cross-helicity, and residual helicity, respectively (see Eq.~\ref{eq:emf}),
under the assumption of comparable time scales $\tau_{\alpha}$, $\tau_{\beta}$,
and $\tau_{\gamma}$ (Eq.~\ref{eq:coeffs}). The main contributions to $\vec{E}_M$
come from the terms related to turbulent energy and residual helicity. The term
related to cross-helicity is nearly negligible, probably due to the small mean
vorticity $\vec{\Omega}$. More interestingly, in the low plasma-$\beta$ model
seen in the left column of Figure~\ref{fig:profiles}g, the term corresponding to
$H$ (green) seems to be a mirror reflection of the term corresponding to $K$
around $y=0$, but somewhat larger in the amplitude. A similar mirror reflection
is observed in terms of the Z components (Fig.~\ref{fig:profiles}h, left),
however, with $-K J_z$ being now somewhat stronger than $H B_z$. In the case of
models with larger plasma-$\beta$ (the middle and right column), the
contribution from the residual helicity H continues to be important, especially
for X components. For Z components, the term related to the turbulent energy K
dominates. The two last rows of Figure~\ref{fig:profiles} surprisingly
demonstrate that the residual helicity might be an important factor of
generation of the X component of electromotive force $\vec{E}_M$. Moreover, its
contribution to the generation of $E_{M,z}$ can be controlled by the strength of
the guide field $B_z$, as seen in the low plasma-$\beta$ regime.


\section{Discussion and Conclusions}
\label{sec:discussion}

The stochastic reconnection is able to generate turbulence and therefore
electromotive force $\vec{E}_M$.  In order to understand how $\vec{E}_M$ is
generated and how it affects the mechanism responsible for its generation, we
analyzed the evolution of turbulent energy $K$, cross-helicity $W$ and residual
helicity $H$, as well as, the production and dissipation of $K$ and $W$, in
direct numerical simulations of the stochastic reconnection for low and mild
plasma-$\beta$ plasmas and Lundquist numbers $S=10^4$ and $10^5$. Moreover, we
also analyzed the distributions of the terms related to the transport
coefficients, $\bar{\alpha}$, $\bar{\beta}$, and $\bar{\gamma}$ expressed by
Eq.~(\ref{eq:coeffs}).

The present study aims to interpret our previous results  of turbulent
reconnection \citep{Kowal_etal:2017, Kowal_etal:2020} in the context of
mean-field theory proposed by \cite{Yokoi_etal:2013} and
\cite{Higashimori_etal:2013}. Differently to their work, the mean and
fluctuating fields are obtained from direct numeric simulations, and not by
means of the mean-ﬁeld equations. As a consequence, the turbulence is
intrinsically developed in our model, i.e. as a natural consequence of the
evolution of reconnection region. \cite{Yokoi_etal:2013} and
\cite{Higashimori_etal:2013} analyzed the importance of $\bar{\beta}$ and
$\bar{\gamma}$ effects. Here we additionally studied the development of
$\bar{\alpha}$ term related to the residual helicity $H$.

Previous studies of turbulent magnetic reconnection in the context of mean-field
theory \cite[e.g.][]{Higashimori_etal:2013, Widmer_etal:2016a, Widmer_etal:2019}
have been done under the assumption of the scale separation between the
reconnection process and developed turbulence, i.e., it was assumed that the
turbulent cascade is developed at scales smaller than the current sheet
thickness. This assumption allows to study the effects of turbulence on the
global magnetic reconnection using two approaches: 1) the set of transport
coefficients $\bar{\alpha}$, $\bar{\beta}$, and $\bar{\gamma}$ incorporated
directly into the MHD equations governing the mean field evolution, or 2) by
controlling the turbulence time scales $\tau_{\alpha}$, $\tau_{\beta}$, and
$\tau_{\gamma}$ and modeling the evolution of turbulent energy $K$,
cross-helicity $W$, and dissipation rate $\varepsilon$ \cite[see,
e.g.,][]{Widmer_etal:2019}. A more advanced approach was considered in
\cite{Widmer_etal:2016b} by applying the sub-grid scale filtering procedure to
high resolution simulations of plasmoid instability, still with turbulence
assumed to operate at scales smaller than the development of the instability.
Apart from the assumed scale separation between the reconnection and turbulence,
however, all these works were limited to 2D geometry, a significant constraint
both for the evolution of stochastic reconnection, as well as, the magnetized
turbulence \cite[see][and references therein, for validation of 3D geometry in
the studies of these phenomena]{Lazarian_etal:2020}.

Our studies are based on the principle that the turbulent cascade can be
developed by the stochastic magnetic reconnection at scales at comparable
scales, extending from the system size down to scales below the current sheet
thickness \cite[see][]{Kowal_etal:2017}. Therefore, at scales $l \gg \delta$,
turbulent eddies may be strong enough to directly affect the structure of the
current sheet. On the other hand, magnetic reconnection is an intrinsic part of
the turbulent cascade \citep[see, e.g.,][]{Eyink_etal:2013, Eyink:2015,
JafariVishniac:2019, Lazarian_etal:2020}. Therefore, the only way of separation
between these two phenomena, from our point of view, is by considering the
profiles of the mean magnetic field and velocity, which solely depend on the $y$
coordinate. This is justified by the fact that, far from the turbulent region,
the plasma is essentially not affected by turbulence, except for the development
of the global inflow transporting the magnetic flux toward the turbulent
reconnection region.

Clearly, the picture described above is not completely compatible with the
mean-field theory of \cite{Yokoi_etal:2013} applied to magnetic reconnection. In
the light of this theory, the cross-helicity is expected to be concentrated near
the individual current sheets with a quadruple distribution across the X-points
\cite[see, e.g.,][]{Widmer_etal:2016b}. A simple averaging over these structures
may result in the cancellation between positive and negative cross-helicity
structures. As explain above, the reconnection happens at all scales of the
turbulent cascade, therefore we rather expect a different manifestation of the
reconnection than a distribution of X-points. Even by assuming the existence of
local individual reconnection events occurring at different scales, their
orientations would become more stochastic once going to smaller and smaller
scales, and resulting in no clear way of avoiding the mentioned cancellation.
Another important point is that the cascade goes through all scales available
over the computational domain, therefore there is no clear indication at what
scale the averaging should be done. As pointed out in \S\ref{sec:averaging}, the
applied decomposition makes the mean field evolution consistent, additionally
justifying our approach. Considering the above points, this work is not aimed to
validate or invalidate the mean-field theory by \cite{Yokoi_etal:2013}, which
was tested in numerical setups significantly different than ours, but rather to
apply the same tools to our direct numerical simulations of stochastic magnetic
reconnection in order to study how the well defined quantities, such as
turbulent energy or cross-helicity, affect the reconnection process in the
presence of turbulence near the global current sheet.

Our aim was to verify any possible correlation between the estimated global
reconnection rate $V_{rec}$ and turbulent energy $K$, cross-helicity $W$, or the
relative helicity $H$, responsible for the $\bar{\beta}$, $\bar{\gamma}$, and
$\bar{\alpha}$ effects, respectively. We should remark, that the $\bar{\alpha}$
effect might be important here due to the presence of a guide field in our
simulations. Also, we should stress that the development of the cross-helicity
is possible since the initial velocity perturbations provide non-zero
large-scale vorticity, although relatively small. Our more detailed analysis of
the term contribution to $P_W$ (Eq.~\ref{eq:pw}), which was not presented here,
indicates the second term related to the Reynolds tensor and magnetic field
gradient as the dominating mechanism of the cross-helicity production.

We conclude our studies with the following remarks:
\begin{enumerate}

\item The turbulent energy $K$ reaches a stationary state within a few Alfvén
times for sufficiently large Lundquist numbers. In this state, it is dominated
by the magnetic turbulent energy $K_{b'}$, being nearly two orders of magnitude
larger than the kinetic one $K_{u'}$ in the case of low plasma-$\beta$ model,
while in the model with plasma-$\beta = 2$ $K_{u'}$ becomes nearly as strong as
$K_{b'}$. The turbulent energies are concentrated near the global current sheet.
The turbulent energy dissipation $\varepsilon_K$ is less efficient comparing to
the production $P_K$ in all cases.

\item The cross-helicity is produced in all models, oscillating around zero with
irregular periods, with the maximum absolute values reaching the order of
magnitude of around $10^{-3}$. The periods of its increase and decay are
strongly correlated with the sign of the cross-helicity production $P_W$. The
dissipation of cross-helicity $\varepsilon_W$ is significantly weaker comparing
to $P_W$ for models characterized by higher Lundquist numbers.

\item The effect of cross-helicity on the reconnection rate seems to be
negligible. Our analysis shows no clear correlation between these two
quantities. Comparing $S=10^5$ models with different plasma-$\beta$, $|W|/K$
saturates rapidly (in time much shorter than $t_A$), and the saturated value
increases with the value of plasma-$\beta$ in the range between $10^{-4}$ and
$10^{-3}$. Since the reconnection rate increases in these models once the
turbulence is developed, this indicates that the requirement of a sufficiently
large $|W|/K$ is not a necessary condition for fast reconnection to take place.
Further studies with different spatial separation procedures, which allow for
determination of the actual profiles of cross-helicity across the XZ planes and
the importance of these profiles on the reconnection rate, might be necessary.
In the case of strong diffusivity (model C) the reconnection rate decays, while
the mean $|W|/K$ grows until the saturation phase, indicating that large values
of Ohmic resistivity and viscosity significantly suppress the way turbulence
affects the reconnection, most probably by dumping the fluctuations at the
length-scales comparable to the current sheet thickness.

\item In all models the X components of terms contributing to the electromotive
force are anti-symmetric and the Z ones are symmetric across the mid-plane, with
the term $W \vec{\Omega}$ being much smaller as compared to other terms. Under
the assumption of comparable turbulent time scales $\tau_\alpha$, $\tau_\beta$,
and $\tau_\gamma$, it indicates that the $\bar{\gamma}$ related terms in
Eq.~\ref{eq:emf} are nearly negligible and do not contribute to the mean
magnetic field evolution. The main contribution would come from $\bar{\alpha}$
\cite[not analyzed in ][]{Yokoi_etal:2013} and $\bar{\beta}$ related terms.
These two terms seem to counter balance each other, especially in the low
plasma-$\beta$ case, since their strengths are comparable. Still, we see that
this model is characterized by the highest reconnection rate $V_{rec}$.
Moreover, we do not observe the loci effect due to the electromotive terms
balance discussed in \cite{Yokoi_etal:2013}. This might be due to the relatively
homogeneous turbulence produced by stochastic reconnection and insufficient
strength of the large-scale vorticity.

\item A possible candidate for explaining the enhancement of the reconnection
rate observed in the presented models is the effect of turbulent magnetic
diffusion. However, even though it is relatively strong in our models, the
residual helicity works against it. Especially, in the low plasma-beta regime,
the regime where the largest increase of reconnection rate is observed. The
balance between these two effects together with the estimation of turbulent
timescales should be addressed in our future studies.
\end{enumerate}

\acknowledgments
{N.N. acknowledges support by the Polish National Science Centre grant No.
2014/15/N/ST9/04622. G.K. acknowledges support from the Brazilian National
Council for Scientific and Technological Development (CNPq no. 304891/2016-9)
and FAPESP (grants 2013/10559-5 and 2019/03301-8). D.F.G. thanks the Brazilian
agencies CNPq (no. 311128/2017-3) and FAPESP (no. 2013/10559-5) for financial
support.  This work has made use of the computing facilities: the Academic
Supercomputing Center in Kraków, Poland (Supercomputer Prometheus/ACK CYFRONET
AGH) and the cluster of Núcleo de Astrofísica Teórica, Universidade Cruzeiro do
Sul, Brazil.}


\bibliography{ms}

\begin{thebibliography}{55}%
\makeatletter
\providecommand \@ifxundefined [1]{%
 \@ifx{#1\undefined}
}%
\providecommand \@ifnum [1]{%
 \ifnum #1\expandafter \@firstoftwo
 \else \expandafter \@secondoftwo
 \fi
}%
\providecommand \@ifx [1]{%
 \ifx #1\expandafter \@firstoftwo
 \else \expandafter \@secondoftwo
 \fi
}%
\providecommand \natexlab [1]{#1}%
\providecommand \enquote  [1]{``#1''}%
\providecommand \bibnamefont  [1]{#1}%
\providecommand \bibfnamefont [1]{#1}%
\providecommand \citenamefont [1]{#1}%
\providecommand \href@noop [0]{\@secondoftwo}%
\providecommand \href [0]{\begingroup \@sanitize@url \@href}%
\providecommand \@href[1]{\@@startlink{#1}\@@href}%
\providecommand \@@href[1]{\endgroup#1\@@endlink}%
\providecommand \@sanitize@url [0]{\catcode `\\12\catcode `\$12\catcode
  `\&12\catcode `\#12\catcode `\^12\catcode `\_12\catcode `\%12\relax}%
\providecommand \@@startlink[1]{}%
\providecommand \@@endlink[0]{}%
\providecommand \url  [0]{\begingroup\@sanitize@url \@url }%
\providecommand \@url [1]{\endgroup\@href {#1}{\urlprefix }}%
\providecommand \urlprefix  [0]{URL }%
\providecommand \Eprint [0]{\href }%
\providecommand \doibase [0]{http://dx.doi.org/}%
\providecommand \selectlanguage [0]{\@gobble}%
\providecommand \bibinfo  [0]{\@secondoftwo}%
\providecommand \bibfield  [0]{\@secondoftwo}%
\providecommand \translation [1]{[#1]}%
\providecommand \BibitemOpen [0]{}%
\providecommand \bibitemStop [0]{}%
\providecommand \bibitemNoStop [0]{.\EOS\space}%
\providecommand \EOS [0]{\spacefactor3000\relax}%
\providecommand \BibitemShut  [1]{\csname bibitem#1\endcsname}%
\let\auto@bib@innerbib\@empty
\bibitem [{\citenamefont {Ahn}\ and\ \citenamefont {Lee}(2020)}]{AhnLee:2020}%
  \BibitemOpen
  \bibfield  {author} {\bibinfo {author} {\bibnamefont {Ahn}, \bibfnamefont
  {M.-H.}}\ and\ \bibinfo {author} {\bibnamefont {Lee}, \bibfnamefont
  {D.-J.}},\ }\bibfield  {title} {\enquote {\bibinfo {title} {{Modified
  Monotonicity Preserving Constraints for High-Resolution Optimized Compact
  Scheme}},}\ }\href {\doibase 10.1007/s10915-020-01221-0} {\bibfield
  {journal} {\bibinfo  {journal} {Journal of Scientific Computing}\ }\textbf
  {\bibinfo {volume} {83}},\ \bibinfo {pages} {34} (\bibinfo {year}
  {2020})}\BibitemShut {NoStop}%
\bibitem [{\citenamefont {{Armstrong}}, \citenamefont {{Rickett}},\ and\
  \citenamefont {{Spangler}}(1995)}]{Armstrong_etal:1995}%
  \BibitemOpen
  \bibfield  {author} {\bibinfo {author} {\bibnamefont {{Armstrong}},
  \bibfnamefont {J.~W.}}, \bibinfo {author} {\bibnamefont {{Rickett}},
  \bibfnamefont {B.~J.}}, \ and\ \bibinfo {author} {\bibnamefont {{Spangler}},
  \bibfnamefont {S.~R.}},\ }\bibfield  {title} {\enquote {\bibinfo {title}
  {{Electron density power spectrum in the local interstellar medium}},}\
  }\href {\doibase 10.1086/175515} {\bibfield  {journal} {\bibinfo  {journal}
  {The Astrophysical Journal}\ }\textbf {\bibinfo {volume} {443}},\ \bibinfo
  {pages} {209--221} (\bibinfo {year} {1995})}\BibitemShut {NoStop}%
\bibitem [{\citenamefont {{Bessho}}\ \emph {et~al.}(2020)\citenamefont
  {{Bessho}}, \citenamefont {{Chen}}, \citenamefont {{Wang}}, \citenamefont
  {{Hesse}}, \citenamefont {{Wilson}},\ and\ \citenamefont
  {{Ng}}}]{Bessho_etal:2020}%
  \BibitemOpen
  \bibfield  {author} {\bibinfo {author} {\bibnamefont {{Bessho}},
  \bibfnamefont {N.}}, \bibinfo {author} {\bibnamefont {{Chen}}, \bibfnamefont
  {L.~J.}}, \bibinfo {author} {\bibnamefont {{Wang}}, \bibfnamefont {S.}},
  \bibinfo {author} {\bibnamefont {{Hesse}}, \bibfnamefont {M.}}, \bibinfo
  {author} {\bibnamefont {{Wilson}}, \bibfnamefont {L.~B., I.}}, \ and\
  \bibinfo {author} {\bibnamefont {{Ng}}, \bibfnamefont {J.}},\ }\bibfield
  {title} {\enquote {\bibinfo {title} {{Magnetic reconnection and kinetic waves
  generated in the Earth's quasi-parallel bow shock}},}\ }\href {\doibase
  10.1063/5.0012443} {\bibfield  {journal} {\bibinfo  {journal} {Physics of
  Plasmas}\ }\textbf {\bibinfo {volume} {27}},\ \bibinfo {pages} {092901}
  (\bibinfo {year} {2020})}\BibitemShut {NoStop}%
\bibitem [{\citenamefont {{Birn}}\ \emph {et~al.}(2001)\citenamefont {{Birn}},
  \citenamefont {{Drake}}, \citenamefont {{Shay}}, \citenamefont {{Rogers}},
  \citenamefont {{Denton}}, \citenamefont {{Hesse}}, \citenamefont
  {{Kuznetsova}}, \citenamefont {{Ma}}, \citenamefont {{Bhattacharjee}},
  \citenamefont {{Otto}},\ and\ \citenamefont {{Pritchett}}}]{Birn_etal:2001}%
  \BibitemOpen
  \bibfield  {author} {\bibinfo {author} {\bibnamefont {{Birn}}, \bibfnamefont
  {J.}}, \bibinfo {author} {\bibnamefont {{Drake}}, \bibfnamefont {J.~F.}},
  \bibinfo {author} {\bibnamefont {{Shay}}, \bibfnamefont {M.~A.}}, \bibinfo
  {author} {\bibnamefont {{Rogers}}, \bibfnamefont {B.~N.}}, \bibinfo {author}
  {\bibnamefont {{Denton}}, \bibfnamefont {R.~E.}}, \bibinfo {author}
  {\bibnamefont {{Hesse}}, \bibfnamefont {M.}}, \bibinfo {author} {\bibnamefont
  {{Kuznetsova}}, \bibfnamefont {M.}}, \bibinfo {author} {\bibnamefont {{Ma}},
  \bibfnamefont {Z.~W.}}, \bibinfo {author} {\bibnamefont {{Bhattacharjee}},
  \bibfnamefont {A.}}, \bibinfo {author} {\bibnamefont {{Otto}}, \bibfnamefont
  {A.}}, \ and\ \bibinfo {author} {\bibnamefont {{Pritchett}}, \bibfnamefont
  {P.~L.}},\ }\bibfield  {title} {\enquote {\bibinfo {title} {{Geospace
  Environmental Modeling (GEM) magnetic reconnection challenge}},}\ }\href
  {\doibase 10.1029/1999JA900449} {\bibfield  {journal} {\bibinfo  {journal}
  {Journal of Geophysical Research}\ }\textbf {\bibinfo {volume} {106}},\
  \bibinfo {pages} {3715--3720} (\bibinfo {year} {2001})}\BibitemShut {NoStop}%
\bibitem [{\citenamefont {{Brandenburg}}(2018)}]{Brandenburg:2018}%
  \BibitemOpen
  \bibfield  {author} {\bibinfo {author} {\bibnamefont {{Brandenburg}},
  \bibfnamefont {A.}},\ }\bibfield  {title} {\enquote {\bibinfo {title}
  {{Advances in mean-field dynamo theory and applications to astrophysical
  turbulence}},}\ }\href {\doibase 10.1017/S0022377818000806} {\bibfield
  {journal} {\bibinfo  {journal} {Journal of Plasma Physics}\ }\textbf
  {\bibinfo {volume} {84}},\ \bibinfo {eid} {735840404} (\bibinfo {year}
  {2018})},\ \Eprint {http://arxiv.org/abs/1801.05384} {arXiv:1801.05384
  [physics.flu-dyn]} \BibitemShut {NoStop}%
\bibitem [{\citenamefont {{Brandenburg}}\ and\ \citenamefont
  {{Subramanian}}(2005)}]{BrandenburgSubramanian:2005}%
  \BibitemOpen
  \bibfield  {author} {\bibinfo {author} {\bibnamefont {{Brandenburg}},
  \bibfnamefont {A.}}\ and\ \bibinfo {author} {\bibnamefont {{Subramanian}},
  \bibfnamefont {K.}},\ }\bibfield  {title} {\enquote {\bibinfo {title}
  {{Astrophysical magnetic fields and nonlinear dynamo theory}},}\ }\href
  {\doibase 10.1016/j.physrep.2005.06.005} {\bibfield  {journal} {\bibinfo
  {journal} {Physics Reports}\ }\textbf {\bibinfo {volume} {417}},\ \bibinfo
  {pages} {1--209} (\bibinfo {year} {2005})},\ \Eprint
  {http://arxiv.org/abs/astro-ph/0405052} {arXiv:astro-ph/0405052 [astro-ph]}
  \BibitemShut {NoStop}%
\bibitem [{\citenamefont {{Cassak}}, \citenamefont {{Shay}},\ and\
  \citenamefont {{Drake}}(2005)}]{Cassak_etal:2005}%
  \BibitemOpen
  \bibfield  {author} {\bibinfo {author} {\bibnamefont {{Cassak}},
  \bibfnamefont {P.~A.}}, \bibinfo {author} {\bibnamefont {{Shay}},
  \bibfnamefont {M.~A.}}, \ and\ \bibinfo {author} {\bibnamefont {{Drake}},
  \bibfnamefont {J.~F.}},\ }\bibfield  {title} {\enquote {\bibinfo {title}
  {{Catastrophe Model for Fast Magnetic Reconnection Onset}},}\ }\href
  {\doibase 10.1103/PhysRevLett.95.235002} {\bibfield  {journal} {\bibinfo
  {journal} {Physical Review Letters}\ }\textbf {\bibinfo {volume} {95}},\
  \bibinfo {eid} {235002} (\bibinfo {year} {2005})},\ \Eprint
  {http://arxiv.org/abs/physics/0502001} {arXiv:physics/0502001
  [physics.plasm-ph]} \BibitemShut {NoStop}%
\bibitem [{\citenamefont {{Charbonneau}}(2014)}]{Charbonneau:2014}%
  \BibitemOpen
  \bibfield  {author} {\bibinfo {author} {\bibnamefont {{Charbonneau}},
  \bibfnamefont {P.}},\ }\bibfield  {title} {\enquote {\bibinfo {title} {{Solar
  Dynamo Theory}},}\ }\href {\doibase 10.1146/annurev-astro-081913-040012}
  {\bibfield  {journal} {\bibinfo  {journal} {Annual Review of Astronomy And
  astrophysics}\ }\textbf {\bibinfo {volume} {52}},\ \bibinfo {pages}
  {251--290} (\bibinfo {year} {2014})}\BibitemShut {NoStop}%
\bibitem [{\citenamefont {{Chepurnov}}\ and\ \citenamefont
  {{Lazarian}}(2010)}]{ChepurnovLazarian:2010}%
  \BibitemOpen
  \bibfield  {author} {\bibinfo {author} {\bibnamefont {{Chepurnov}},
  \bibfnamefont {A.}}\ and\ \bibinfo {author} {\bibnamefont {{Lazarian}},
  \bibfnamefont {A.}},\ }\bibfield  {title} {\enquote {\bibinfo {title}
  {{Extending the Big Power Law in the Sky with Turbulence Spectra from
  Wisconsin H{$\alpha$} Mapper Data}},}\ }\href {\doibase
  10.1088/0004-637X/710/1/853} {\bibfield  {journal} {\bibinfo  {journal} {The
  Astrophysical Journal}\ }\textbf {\bibinfo {volume} {710}},\ \bibinfo {pages}
  {853--858} (\bibinfo {year} {2010})},\ \Eprint
  {http://arxiv.org/abs/0905.4413} {arXiv:0905.4413 [astro-ph.GA]} \BibitemShut
  {NoStop}%
\bibitem [{\citenamefont {{Eyink}}\ \emph {et~al.}(2013)\citenamefont
  {{Eyink}}, \citenamefont {{Vishniac}}, \citenamefont {{Lalescu}},
  \citenamefont {{Aluie}}, \citenamefont {{Kanov}}, \citenamefont
  {{B{\"u}rger}}, \citenamefont {{Burns}}, \citenamefont {{Meneveau}},\ and\
  \citenamefont {{Szalay}}}]{Eyink_etal:2013}%
  \BibitemOpen
  \bibfield  {author} {\bibinfo {author} {\bibnamefont {{Eyink}}, \bibfnamefont
  {G.}}, \bibinfo {author} {\bibnamefont {{Vishniac}}, \bibfnamefont {E.}},
  \bibinfo {author} {\bibnamefont {{Lalescu}}, \bibfnamefont {C.}}, \bibinfo
  {author} {\bibnamefont {{Aluie}}, \bibfnamefont {H.}}, \bibinfo {author}
  {\bibnamefont {{Kanov}}, \bibfnamefont {K.}}, \bibinfo {author} {\bibnamefont
  {{B{\"u}rger}}, \bibfnamefont {K.}}, \bibinfo {author} {\bibnamefont
  {{Burns}}, \bibfnamefont {R.}}, \bibinfo {author} {\bibnamefont {{Meneveau}},
  \bibfnamefont {C.}}, \ and\ \bibinfo {author} {\bibnamefont {{Szalay}},
  \bibfnamefont {A.}},\ }\bibfield  {title} {\enquote {\bibinfo {title}
  {{Flux-freezing breakdown in high-conductivity magnetohydrodynamic
  turbulence}},}\ }\href {\doibase 10.1038/nature12128} {\bibfield  {journal}
  {\bibinfo  {journal} {Nature}\ }\textbf {\bibinfo {volume} {497}},\ \bibinfo
  {pages} {466--469} (\bibinfo {year} {2013})}\BibitemShut {NoStop}%
\bibitem [{\citenamefont {{Eyink}}(2015)}]{Eyink:2015}%
  \BibitemOpen
  \bibfield  {author} {\bibinfo {author} {\bibnamefont {{Eyink}}, \bibfnamefont
  {G.~L.}},\ }\bibfield  {title} {\enquote {\bibinfo {title} {{Turbulent
  General Magnetic Reconnection}},}\ }\href {\doibase
  10.1088/0004-637X/807/2/137} {\bibfield  {journal} {\bibinfo  {journal} {The
  Astrophysical Journal}\ }\textbf {\bibinfo {volume} {807}},\ \bibinfo {eid}
  {137} (\bibinfo {year} {2015})},\ \Eprint {http://arxiv.org/abs/1412.2254}
  {arXiv:1412.2254 [astro-ph.SR]} \BibitemShut {NoStop}%
\bibitem [{\citenamefont {{Falceta-Gon{\c{c}}alves}}\ \emph
  {et~al.}(2015)\citenamefont {{Falceta-Gon{\c{c}}alves}}, \citenamefont
  {{Bonnell}}, \citenamefont {{Kowal}}, \citenamefont {{L{\'e}pine}},\ and\
  \citenamefont {{Braga}}}]{FalcetaGoncalves_etal:2015}%
  \BibitemOpen
  \bibfield  {author} {\bibinfo {author} {\bibnamefont
  {{Falceta-Gon{\c{c}}alves}}, \bibfnamefont {D.}}, \bibinfo {author}
  {\bibnamefont {{Bonnell}}, \bibfnamefont {I.}}, \bibinfo {author}
  {\bibnamefont {{Kowal}}, \bibfnamefont {G.}}, \bibinfo {author} {\bibnamefont
  {{L{\'e}pine}}, \bibfnamefont {J.~R.~D.}}, \ and\ \bibinfo {author}
  {\bibnamefont {{Braga}}, \bibfnamefont {C.~A.~S.}},\ }\bibfield  {title}
  {\enquote {\bibinfo {title} {{The onset of large-scale turbulence in the
  interstellar medium of spiral galaxies}},}\ }\href {\doibase
  10.1093/mnras/stu2127} {\bibfield  {journal} {\bibinfo  {journal} {Monthly
  Notices of the Royal Astronomical Society}\ }\textbf {\bibinfo {volume}
  {446}},\ \bibinfo {pages} {973--989} (\bibinfo {year} {2015})},\ \Eprint
  {http://arxiv.org/abs/1410.2774} {arXiv:1410.2774 [astro-ph.GA]} \BibitemShut
  {NoStop}%
\bibitem [{\citenamefont {{Gonzalez}}\ and\ \citenamefont
  {{Parker}}(2016)}]{GonzalezParker:2016}%
  \BibitemOpen
  \bibfield  {author} {\bibinfo {author} {\bibnamefont {{Gonzalez}},
  \bibfnamefont {W.}}\ and\ \bibinfo {author} {\bibnamefont {{Parker}},
  \bibfnamefont {E.}},\ }\href {\doibase 10.1007/978-3-319-26432-5} {\emph
  {\bibinfo {title} {{Magnetic Reconnection}}}},\ Vol.\ \bibinfo {volume}
  {427}\ (\bibinfo  {publisher} {Springer},\ \bibinfo {year}
  {2016})\BibitemShut {NoStop}%
\bibitem [{\citenamefont {Gottlieb}, \citenamefont {Ketcheson},\ and\
  \citenamefont {Shu}(2011)}]{Gottlieb_etal:2011}%
  \BibitemOpen
  \bibfield  {author} {\bibinfo {author} {\bibnamefont {Gottlieb},
  \bibfnamefont {S.}}, \bibinfo {author} {\bibnamefont {Ketcheson},
  \bibfnamefont {D.}}, \ and\ \bibinfo {author} {\bibnamefont {Shu},
  \bibfnamefont {C.-W.}},\ }\href {\doibase 10.1142/7498} {\emph {\bibinfo
  {title} {Strong Stability Preserving Runge-Kutta and Multistep Time
  Discretizations}}}\ (\bibinfo  {publisher} {WORLD SCIENTIFIC},\ \bibinfo
  {year} {2011})\ \Eprint
  {http://arxiv.org/abs/https://www.worldscientific.com/doi/pdf/10.1142/7498}
  {https://www.worldscientific.com/doi/pdf/10.1142/7498} \BibitemShut {NoStop}%
\bibitem [{\citenamefont {{Higashimori}}, \citenamefont {{Yokoi}},\ and\
  \citenamefont {{Hoshino}}(2013)}]{Higashimori_etal:2013}%
  \BibitemOpen
  \bibfield  {author} {\bibinfo {author} {\bibnamefont {{Higashimori}},
  \bibfnamefont {K.}}, \bibinfo {author} {\bibnamefont {{Yokoi}}, \bibfnamefont
  {N.}}, \ and\ \bibinfo {author} {\bibnamefont {{Hoshino}}, \bibfnamefont
  {M.}},\ }\bibfield  {title} {\enquote {\bibinfo {title} {{Explosive Turbulent
  Magnetic Reconnection}},}\ }\href {\doibase 10.1103/PhysRevLett.110.255001}
  {\bibfield  {journal} {\bibinfo  {journal} {Physical Review Letters}\
  }\textbf {\bibinfo {volume} {110}},\ \bibinfo {eid} {255001} (\bibinfo {year}
  {2013})},\ \Eprint {http://arxiv.org/abs/1305.6695} {arXiv:1305.6695
  [astro-ph.EP]} \BibitemShut {NoStop}%
\bibitem [{\citenamefont {{Hughes}}\ and\ \citenamefont
  {{Tobias}}(2010)}]{HughesTobias:2010}%
  \BibitemOpen
  \bibfield  {author} {\bibinfo {author} {\bibnamefont {{Hughes}},
  \bibfnamefont {D.~W.}}\ and\ \bibinfo {author} {\bibnamefont {{Tobias}},
  \bibfnamefont {S.~M.}},\ }\enquote {\bibinfo {title} {{An Introduction to
  Mean Field Dynamo Theory}},}\ in\ \href {\doibase 10.1142/9789814291552_0002}
  {\emph {\bibinfo {booktitle} {Relaxation Dynamics in Laboratory and
  Astrophysical Plasmas. Edited by DIAMOND PATRICK H ET AL. Published by World
  Scientific Publishing Co. Pte. Ltd}}}\ (\bibinfo  {publisher} {World
  Scientific Publishing Company},\ \bibinfo {year} {2010})\ pp.\ \bibinfo
  {pages} {15--48}\BibitemShut {NoStop}%
\bibitem [{\citenamefont {{Jabbari}}\ \emph {et~al.}(2016)\citenamefont
  {{Jabbari}}, \citenamefont {{Brandenburg}}, \citenamefont {{Mitra}},
  \citenamefont {{Kleeorin}},\ and\ \citenamefont
  {{Rogachevskii}}}]{Jabbari_etal:2016}%
  \BibitemOpen
  \bibfield  {author} {\bibinfo {author} {\bibnamefont {{Jabbari}},
  \bibfnamefont {S.}}, \bibinfo {author} {\bibnamefont {{Brandenburg}},
  \bibfnamefont {A.}}, \bibinfo {author} {\bibnamefont {{Mitra}}, \bibfnamefont
  {D.}}, \bibinfo {author} {\bibnamefont {{Kleeorin}}, \bibfnamefont {N.}}, \
  and\ \bibinfo {author} {\bibnamefont {{Rogachevskii}}, \bibfnamefont {I.}},\
  }\bibfield  {title} {\enquote {\bibinfo {title} {{Turbulent reconnection of
  magnetic bipoles in stratified turbulence}},}\ }\href {\doibase
  10.1093/mnras/stw888} {\bibfield  {journal} {\bibinfo  {journal} {Monthly
  Notices of the Royal Astronomical Society}\ }\textbf {\bibinfo {volume}
  {459}},\ \bibinfo {pages} {4046--4056} (\bibinfo {year} {2016})},\ \Eprint
  {http://arxiv.org/abs/1601.08167} {arXiv:1601.08167 [astro-ph.SR]}
  \BibitemShut {NoStop}%
\bibitem [{\citenamefont {{Jafari}}\ and\ \citenamefont
  {{Vishniac}}(2019)}]{JafariVishniac:2019}%
  \BibitemOpen
  \bibfield  {author} {\bibinfo {author} {\bibnamefont {{Jafari}},
  \bibfnamefont {A.}}\ and\ \bibinfo {author} {\bibnamefont {{Vishniac}},
  \bibfnamefont {E.}},\ }\bibfield  {title} {\enquote {\bibinfo {title}
  {{Topology and stochasticity of turbulent magnetic fields}},}\ }\href
  {\doibase 10.1103/PhysRevE.100.013201} {\bibfield  {journal} {\bibinfo
  {journal} {Physical Review E}\ }\textbf {\bibinfo {volume} {100}},\ \bibinfo
  {eid} {013201} (\bibinfo {year} {2019})}\BibitemShut {NoStop}%
\bibitem [{\citenamefont {{Kowal}}\ \emph {et~al.}(2017)\citenamefont
  {{Kowal}}, \citenamefont {{Falceta-Gon{\c c}alves}}, \citenamefont
  {{Lazarian}},\ and\ \citenamefont {{Vishniac}}}]{Kowal_etal:2017}%
  \BibitemOpen
  \bibfield  {author} {\bibinfo {author} {\bibnamefont {{Kowal}}, \bibfnamefont
  {G.}}, \bibinfo {author} {\bibnamefont {{Falceta-Gon{\c c}alves}},
  \bibfnamefont {D.~A.}}, \bibinfo {author} {\bibnamefont {{Lazarian}},
  \bibfnamefont {A.}}, \ and\ \bibinfo {author} {\bibnamefont {{Vishniac}},
  \bibfnamefont {E.~T.}},\ }\bibfield  {title} {\enquote {\bibinfo {title}
  {{Statistics of Reconnection-driven Turbulence}},}\ }\href {\doibase
  10.3847/1538-4357/aa6001} {\bibfield  {journal} {\bibinfo  {journal} {The
  Astrophysical Journal}\ }\textbf {\bibinfo {volume} {838}},\ \bibinfo {eid}
  {91} (\bibinfo {year} {2017})},\ \Eprint {http://arxiv.org/abs/1611.03914}
  {arXiv:1611.03914} \BibitemShut {NoStop}%
\bibitem [{\citenamefont {{Kowal}}\ \emph {et~al.}(2020)\citenamefont
  {{Kowal}}, \citenamefont {{Falceta-Gon{\c{c}}alves}}, \citenamefont
  {{Lazarian}},\ and\ \citenamefont {{Vishniac}}}]{Kowal_etal:2020}%
  \BibitemOpen
  \bibfield  {author} {\bibinfo {author} {\bibnamefont {{Kowal}}, \bibfnamefont
  {G.}}, \bibinfo {author} {\bibnamefont {{Falceta-Gon{\c{c}}alves}},
  \bibfnamefont {D.~A.}}, \bibinfo {author} {\bibnamefont {{Lazarian}},
  \bibfnamefont {A.}}, \ and\ \bibinfo {author} {\bibnamefont {{Vishniac}},
  \bibfnamefont {E.~T.}},\ }\bibfield  {title} {\enquote {\bibinfo {title}
  {{Kelvin-Helmholtz versus Tearing Instability: What Drives Turbulence in
  Stochastic Reconnection?}}}\ }\href {\doibase 10.3847/1538-4357/ab7a13}
  {\bibfield  {journal} {\bibinfo  {journal} {The Astrophysical Journal}\
  }\textbf {\bibinfo {volume} {892}},\ \bibinfo {eid} {50} (\bibinfo {year}
  {2020})},\ \Eprint {http://arxiv.org/abs/1909.09179} {arXiv:1909.09179
  [astro-ph.HE]} \BibitemShut {NoStop}%
\bibitem [{\citenamefont {{Kowal}}\ \emph {et~al.}(2009)\citenamefont
  {{Kowal}}, \citenamefont {{Lazarian}}, \citenamefont {{Vishniac}},\ and\
  \citenamefont {{Otmianowska-Mazur}}}]{Kowal_etal:2009}%
  \BibitemOpen
  \bibfield  {author} {\bibinfo {author} {\bibnamefont {{Kowal}}, \bibfnamefont
  {G.}}, \bibinfo {author} {\bibnamefont {{Lazarian}}, \bibfnamefont {A.}},
  \bibinfo {author} {\bibnamefont {{Vishniac}}, \bibfnamefont {E.~T.}}, \ and\
  \bibinfo {author} {\bibnamefont {{Otmianowska-Mazur}}, \bibfnamefont {K.}},\
  }\bibfield  {title} {\enquote {\bibinfo {title} {{Numerical Tests of Fast
  Reconnection in Weakly Stochastic Magnetic Fields}},}\ }\href {\doibase
  10.1088/0004-637X/700/1/63} {\bibfield  {journal} {\bibinfo  {journal} {The
  Astrophysical Journal}\ }\textbf {\bibinfo {volume} {700}},\ \bibinfo {pages}
  {63--85} (\bibinfo {year} {2009})},\ \Eprint {http://arxiv.org/abs/0903.2052}
  {arXiv:0903.2052 [astro-ph.GA]} \BibitemShut {NoStop}%
\bibitem [{\citenamefont {{Kowal}}\ \emph {et~al.}(2012)\citenamefont
  {{Kowal}}, \citenamefont {{Lazarian}}, \citenamefont {{Vishniac}},\ and\
  \citenamefont {{Otmianowska-Mazur}}}]{Kowal_etal:2012}%
  \BibitemOpen
  \bibfield  {author} {\bibinfo {author} {\bibnamefont {{Kowal}}, \bibfnamefont
  {G.}}, \bibinfo {author} {\bibnamefont {{Lazarian}}, \bibfnamefont {A.}},
  \bibinfo {author} {\bibnamefont {{Vishniac}}, \bibfnamefont {E.~T.}}, \ and\
  \bibinfo {author} {\bibnamefont {{Otmianowska-Mazur}}, \bibfnamefont {K.}},\
  }\bibfield  {title} {\enquote {\bibinfo {title} {{Reconnection studies under
  different types of turbulence driving}},}\ }\href {\doibase
  10.5194/npg-19-297-2012} {\bibfield  {journal} {\bibinfo  {journal}
  {Nonlinear Processes in Geophysics}\ }\textbf {\bibinfo {volume} {19}},\
  \bibinfo {pages} {297--314} (\bibinfo {year} {2012})},\ \Eprint
  {http://arxiv.org/abs/1203.2971} {arXiv:1203.2971 [astro-ph.SR]} \BibitemShut
  {NoStop}%
\bibitem [{\citenamefont {{Krause}}\ and\ \citenamefont
  {{Raedler}}(1980)}]{KrauseRaedler:1980}%
  \BibitemOpen
  \bibfield  {author} {\bibinfo {author} {\bibnamefont {{Krause}},
  \bibfnamefont {F.}}\ and\ \bibinfo {author} {\bibnamefont {{Raedler}},
  \bibfnamefont {K.~H.}},\ }\href@noop {} {\emph {\bibinfo {title} {{Mean-field
  magnetohydrodynamics and dynamo theory}}}}\ (\bibinfo  {publisher} {Pergamon
  Press},\ \bibinfo {year} {1980})\BibitemShut {NoStop}%
\bibitem [{\citenamefont {{Lazarian}}\ \emph {et~al.}(2020)\citenamefont
  {{Lazarian}}, \citenamefont {{Eyink}}, \citenamefont {{Jafari}},
  \citenamefont {{Kowal}}, \citenamefont {{Li}}, \citenamefont {{Xu}},\ and\
  \citenamefont {{Vishniac}}}]{Lazarian_etal:2020}%
  \BibitemOpen
  \bibfield  {author} {\bibinfo {author} {\bibnamefont {{Lazarian}},
  \bibfnamefont {A.}}, \bibinfo {author} {\bibnamefont {{Eyink}}, \bibfnamefont
  {G.~L.}}, \bibinfo {author} {\bibnamefont {{Jafari}}, \bibfnamefont {A.}},
  \bibinfo {author} {\bibnamefont {{Kowal}}, \bibfnamefont {G.}}, \bibinfo
  {author} {\bibnamefont {{Li}}, \bibfnamefont {H.}}, \bibinfo {author}
  {\bibnamefont {{Xu}}, \bibfnamefont {S.}}, \ and\ \bibinfo {author}
  {\bibnamefont {{Vishniac}}, \bibfnamefont {E.~T.}},\ }\bibfield  {title}
  {\enquote {\bibinfo {title} {{3D turbulent reconnection: Theory, tests, and
  astrophysical implications}},}\ }\href {\doibase 10.1063/1.5110603}
  {\bibfield  {journal} {\bibinfo  {journal} {Physics of Plasmas}\ }\textbf
  {\bibinfo {volume} {27}},\ \bibinfo {eid} {012305} (\bibinfo {year}
  {2020})},\ \Eprint {http://arxiv.org/abs/2001.00868} {arXiv:2001.00868
  [astro-ph.HE]} \BibitemShut {NoStop}%
\bibitem [{\citenamefont {{Lazarian}}\ and\ \citenamefont
  {{Vishniac}}(1999)}]{LazarianVishniac:1999}%
  \BibitemOpen
  \bibfield  {author} {\bibinfo {author} {\bibnamefont {{Lazarian}},
  \bibfnamefont {A.}}\ and\ \bibinfo {author} {\bibnamefont {{Vishniac}},
  \bibfnamefont {E.~T.}},\ }\bibfield  {title} {\enquote {\bibinfo {title}
  {{Reconnection in a Weakly Stochastic Field}},}\ }\href {\doibase
  10.1086/307233} {\bibfield  {journal} {\bibinfo  {journal} {The Astrophysical
  Journal}\ }\textbf {\bibinfo {volume} {517}},\ \bibinfo {pages} {700--718}
  (\bibinfo {year} {1999})},\ \Eprint {http://arxiv.org/abs/astro-ph/9811037}
  {astro-ph/9811037} \BibitemShut {NoStop}%
\bibitem [{\citenamefont {{Liu}}\ \emph {et~al.}(2017)\citenamefont {{Liu}},
  \citenamefont {{Hesse}}, \citenamefont {{Guo}}, \citenamefont {{Daughton}},
  \citenamefont {{Li}}, \citenamefont {{Cassak}},\ and\ \citenamefont
  {{Shay}}}]{Liu_etal:2017}%
  \BibitemOpen
  \bibfield  {author} {\bibinfo {author} {\bibnamefont {{Liu}}, \bibfnamefont
  {Y.-H.}}, \bibinfo {author} {\bibnamefont {{Hesse}}, \bibfnamefont {M.}},
  \bibinfo {author} {\bibnamefont {{Guo}}, \bibfnamefont {F.}}, \bibinfo
  {author} {\bibnamefont {{Daughton}}, \bibfnamefont {W.}}, \bibinfo {author}
  {\bibnamefont {{Li}}, \bibfnamefont {H.}}, \bibinfo {author} {\bibnamefont
  {{Cassak}}, \bibfnamefont {P.~A.}}, \ and\ \bibinfo {author} {\bibnamefont
  {{Shay}}, \bibfnamefont {M.~A.}},\ }\bibfield  {title} {\enquote {\bibinfo
  {title} {{Why does Steady-State Magnetic Reconnection have a Maximum Local
  Rate of Order 0.1?}}}\ }\href {\doibase 10.1103/PhysRevLett.118.085101}
  {\bibfield  {journal} {\bibinfo  {journal} {Physical Review Letters}\
  }\textbf {\bibinfo {volume} {118}},\ \bibinfo {eid} {085101} (\bibinfo {year}
  {2017})},\ \Eprint {http://arxiv.org/abs/1611.07859} {arXiv:1611.07859
  [physics.plasm-ph]} \BibitemShut {NoStop}%
\bibitem [{\citenamefont {{Markidis}}\ \emph {et~al.}(2014)\citenamefont
  {{Markidis}}, \citenamefont {{Lapenta}}, \citenamefont {{Delzanno}},
  \citenamefont {{Henri}}, \citenamefont {{Goldman}}, \citenamefont {{Newman}},
  \citenamefont {{Intrator}},\ and\ \citenamefont
  {{Laure}}}]{Markidis_etal:2014}%
  \BibitemOpen
  \bibfield  {author} {\bibinfo {author} {\bibnamefont {{Markidis}},
  \bibfnamefont {S.}}, \bibinfo {author} {\bibnamefont {{Lapenta}},
  \bibfnamefont {G.}}, \bibinfo {author} {\bibnamefont {{Delzanno}},
  \bibfnamefont {G.~L.}}, \bibinfo {author} {\bibnamefont {{Henri}},
  \bibfnamefont {P.}}, \bibinfo {author} {\bibnamefont {{Goldman}},
  \bibfnamefont {M.~V.}}, \bibinfo {author} {\bibnamefont {{Newman}},
  \bibfnamefont {D.~L.}}, \bibinfo {author} {\bibnamefont {{Intrator}},
  \bibfnamefont {T.}}, \ and\ \bibinfo {author} {\bibnamefont {{Laure}},
  \bibfnamefont {E.}},\ }\bibfield  {title} {\enquote {\bibinfo {title}
  {{Signatures of secondary collisionless magnetic reconnection driven by kink
  instability of a flux rope}},}\ }\href {\doibase
  10.1088/0741-3335/56/6/064010} {\bibfield  {journal} {\bibinfo  {journal}
  {Plasma Physics and Controlled Fusion}\ }\textbf {\bibinfo {volume} {56}},\
  \bibinfo {eid} {064010} (\bibinfo {year} {2014})},\ \Eprint
  {http://arxiv.org/abs/1408.1144} {arXiv:1408.1144 [physics.plasm-ph]}
  \BibitemShut {NoStop}%
\bibitem [{\citenamefont {{Marsch}}\ and\ \citenamefont
  {{Mangeney}}(1987)}]{MarschMangeney:1987}%
  \BibitemOpen
  \bibfield  {author} {\bibinfo {author} {\bibnamefont {{Marsch}},
  \bibfnamefont {E.}}\ and\ \bibinfo {author} {\bibnamefont {{Mangeney}},
  \bibfnamefont {A.}},\ }\bibfield  {title} {\enquote {\bibinfo {title} {{Ideal
  MHD equations in terms of compressible Els{\"a}sser variables}},}\ }\href
  {\doibase 10.1029/JA092iA07p07363} {\bibfield  {journal} {\bibinfo  {journal}
  {Journal of Geophysical Research}\ }\textbf {\bibinfo {volume} {92}},\
  \bibinfo {pages} {7363--7367} (\bibinfo {year} {1987})}\BibitemShut {NoStop}%
\bibitem [{\citenamefont {{McManus}}\ \emph {et~al.}(2020)\citenamefont
  {{McManus}}, \citenamefont {{Bowen}}, \citenamefont {{Mallet}}, \citenamefont
  {{Chen}}, \citenamefont {{Chandran}}, \citenamefont {{Bale}}, \citenamefont
  {{Larson}}, \citenamefont {{Dudok de Wit}}, \citenamefont {{Kasper}},
  \citenamefont {{Stevens}}, \citenamefont {{Whittlesey}}, \citenamefont
  {{Livi}}, \citenamefont {{Korreck}}, \citenamefont {{Goetz}}, \citenamefont
  {{Harvey}}, \citenamefont {{Pulupa}}, \citenamefont {{MacDowall}},
  \citenamefont {{Malaspina}}, \citenamefont {{Case}},\ and\ \citenamefont
  {{Bonnell}}}]{McManus_etal:2020}%
  \BibitemOpen
  \bibfield  {author} {\bibinfo {author} {\bibnamefont {{McManus}},
  \bibfnamefont {M.~D.}}, \bibinfo {author} {\bibnamefont {{Bowen}},
  \bibfnamefont {T.~A.}}, \bibinfo {author} {\bibnamefont {{Mallet}},
  \bibfnamefont {A.}}, \bibinfo {author} {\bibnamefont {{Chen}}, \bibfnamefont
  {C.~H.~K.}}, \bibinfo {author} {\bibnamefont {{Chandran}}, \bibfnamefont
  {B.~D.~G.}}, \bibinfo {author} {\bibnamefont {{Bale}}, \bibfnamefont
  {S.~D.}}, \bibinfo {author} {\bibnamefont {{Larson}}, \bibfnamefont {D.~E.}},
  \bibinfo {author} {\bibnamefont {{Dudok de Wit}}, \bibfnamefont {T.}},
  \bibinfo {author} {\bibnamefont {{Kasper}}, \bibfnamefont {J.~C.}}, \bibinfo
  {author} {\bibnamefont {{Stevens}}, \bibfnamefont {M.}}, \bibinfo {author}
  {\bibnamefont {{Whittlesey}}, \bibfnamefont {P.}}, \bibinfo {author}
  {\bibnamefont {{Livi}}, \bibfnamefont {R.}}, \bibinfo {author} {\bibnamefont
  {{Korreck}}, \bibfnamefont {K.~E.}}, \bibinfo {author} {\bibnamefont
  {{Goetz}}, \bibfnamefont {K.}}, \bibinfo {author} {\bibnamefont {{Harvey}},
  \bibfnamefont {P.~R.}}, \bibinfo {author} {\bibnamefont {{Pulupa}},
  \bibfnamefont {M.}}, \bibinfo {author} {\bibnamefont {{MacDowall}},
  \bibfnamefont {R.~J.}}, \bibinfo {author} {\bibnamefont {{Malaspina}},
  \bibfnamefont {D.~M.}}, \bibinfo {author} {\bibnamefont {{Case}},
  \bibfnamefont {A.~W.}}, \ and\ \bibinfo {author} {\bibnamefont {{Bonnell}},
  \bibfnamefont {J.~W.}},\ }\bibfield  {title} {\enquote {\bibinfo {title}
  {{Cross Helicity Reversals in Magnetic Switchbacks}},}\ }\href {\doibase
  10.3847/1538-4365/ab6dce} {\bibfield  {journal} {\bibinfo  {journal} {The
  Astrophysical Journal Supplement Series}\ }\textbf {\bibinfo {volume}
  {246}},\ \bibinfo {eid} {67} (\bibinfo {year} {2020})},\ \Eprint
  {http://arxiv.org/abs/1912.07823} {arXiv:1912.07823 [physics.space-ph]}
  \BibitemShut {NoStop}%
\bibitem [{\citenamefont {{Mignone}}(2007)}]{Mignone:2007}%
  \BibitemOpen
  \bibfield  {author} {\bibinfo {author} {\bibnamefont {{Mignone}},
  \bibfnamefont {A.}},\ }\bibfield  {title} {\enquote {\bibinfo {title} {{A
  simple and accurate Riemann solver for isothermal MHD}},}\ }\href {\doibase
  10.1016/j.jcp.2007.01.033} {\bibfield  {journal} {\bibinfo  {journal}
  {Journal of Computational Physics}\ }\textbf {\bibinfo {volume} {225}},\
  \bibinfo {pages} {1427--1441} (\bibinfo {year} {2007})},\ \Eprint
  {http://arxiv.org/abs/astro-ph/0701798} {astro-ph/0701798} \BibitemShut
  {NoStop}%
\bibitem [{\citenamefont {{Moffatt}}(1978)}]{Moffatt:1978}%
  \BibitemOpen
  \bibfield  {author} {\bibinfo {author} {\bibnamefont {{Moffatt}},
  \bibfnamefont {H.~K.}},\ }\href@noop {} {\emph {\bibinfo {title} {{Magnetic
  field generation in electrically conducting fluids}}}}\ (\bibinfo
  {publisher} {Cambridge University Press},\ \bibinfo {year}
  {1978})\BibitemShut {NoStop}%
\bibitem [{\citenamefont {{Nakamura}}\ \emph {et~al.}(2020)\citenamefont
  {{Nakamura}}, \citenamefont {{Stawarz}}, \citenamefont {{Hasegawa}},
  \citenamefont {{Narita}}, \citenamefont {{Franci}}, \citenamefont {{Wilder}},
  \citenamefont {{Nakamura}},\ and\ \citenamefont
  {{Nystrom}}}]{Nakamura_etal:2020}%
  \BibitemOpen
  \bibfield  {author} {\bibinfo {author} {\bibnamefont {{Nakamura}},
  \bibfnamefont {T.~{\^A}. K. {\^A}.~M.}}, \bibinfo {author} {\bibnamefont
  {{Stawarz}}, \bibfnamefont {J.~{\^A}.~E.}}, \bibinfo {author} {\bibnamefont
  {{Hasegawa}}, \bibfnamefont {H.}}, \bibinfo {author} {\bibnamefont
  {{Narita}}, \bibfnamefont {Y.}}, \bibinfo {author} {\bibnamefont {{Franci}},
  \bibfnamefont {L.}}, \bibinfo {author} {\bibnamefont {{Wilder}},
  \bibfnamefont {F.~D.}}, \bibinfo {author} {\bibnamefont {{Nakamura}},
  \bibfnamefont {R.}}, \ and\ \bibinfo {author} {\bibnamefont {{Nystrom}},
  \bibfnamefont {W.~{\^A}.~D.}},\ }\bibfield  {title} {\enquote {\bibinfo
  {title} {{Effects of Fluctuating Magnetic Field on the Growth of the
  Kelvin-Helmholtz Instability at the Earth's Magnetopause}},}\ }\href
  {\doibase 10.1029/2019JA027515} {\bibfield  {journal} {\bibinfo  {journal}
  {Journal of Geophysical Research (Space Physics)}\ }\textbf {\bibinfo
  {volume} {125}},\ \bibinfo {eid} {e27515} (\bibinfo {year}
  {2020})}\BibitemShut {NoStop}%
\bibitem [{\citenamefont {{Padoan}}\ \emph {et~al.}(2009)\citenamefont
  {{Padoan}}, \citenamefont {{Juvela}}, \citenamefont {{Kritsuk}},\ and\
  \citenamefont {{Norman}}}]{Padoan_etal:2009}%
  \BibitemOpen
  \bibfield  {author} {\bibinfo {author} {\bibnamefont {{Padoan}},
  \bibfnamefont {P.}}, \bibinfo {author} {\bibnamefont {{Juvela}},
  \bibfnamefont {M.}}, \bibinfo {author} {\bibnamefont {{Kritsuk}},
  \bibfnamefont {A.}}, \ and\ \bibinfo {author} {\bibnamefont {{Norman}},
  \bibfnamefont {M.~L.}},\ }\bibfield  {title} {\enquote {\bibinfo {title}
  {{The Power Spectrum of Turbulence in NGC 1333: Outflows or Large-Scale
  Driving?}}}\ }\href {\doibase 10.1088/0004-637X/707/2/L153} {\bibfield
  {journal} {\bibinfo  {journal} {The Astrophysical Journall}\ }\textbf
  {\bibinfo {volume} {707}},\ \bibinfo {pages} {L153--L157} (\bibinfo {year}
  {2009})},\ \Eprint {http://arxiv.org/abs/0910.1384} {arXiv:0910.1384}
  \BibitemShut {NoStop}%
\bibitem [{\citenamefont {{Parker}}(1957)}]{Parker:1957}%
  \BibitemOpen
  \bibfield  {author} {\bibinfo {author} {\bibnamefont {{Parker}},
  \bibfnamefont {E.~N.}},\ }\bibfield  {title} {\enquote {\bibinfo {title}
  {{Sweet's Mechanism for Merging Magnetic Fields in Conducting Fluids}},}\
  }\href {\doibase 10.1029/JZ062i004p00509} {\bibfield  {journal} {\bibinfo
  {journal} {Journal of Geophysical Research}\ }\textbf {\bibinfo {volume}
  {62}},\ \bibinfo {pages} {509--520} (\bibinfo {year} {1957})}\BibitemShut
  {NoStop}%
\bibitem [{\citenamefont {{Sagaut}}(2006)}]{Sagaut:2006}%
  \BibitemOpen
  \bibfield  {author} {\bibinfo {author} {\bibnamefont {{Sagaut}},
  \bibfnamefont {P.}},\ }\href {\doibase 10.1007/b137536} {\emph {\bibinfo
  {title} {{Large Eddy Simulation for Incompressible Flows}}}}\ (\bibinfo
  {publisher} {Springer, Berlin, Heidelberg},\ \bibinfo {year}
  {2006})\BibitemShut {NoStop}%
\bibitem [{\citenamefont {{Schrijver}}\ and\ \citenamefont
  {{Siscoe}}(2009)}]{SchrijverSiscoe:2009}%
  \BibitemOpen
  \bibfield  {author} {\bibinfo {author} {\bibnamefont {{Schrijver}},
  \bibfnamefont {C.~J.}}\ and\ \bibinfo {author} {\bibnamefont {{Siscoe}},
  \bibfnamefont {G.~L.}},\ }\href@noop {} {\emph {\bibinfo {title}
  {{Heliophysics: Plasma Physics of the Local Cosmos}}}}\ (\bibinfo
  {publisher} {Cambridge University Press},\ \bibinfo {year}
  {2009})\BibitemShut {NoStop}%
\bibitem [{\citenamefont {{Shay}}\ and\ \citenamefont
  {{Drake}}(1998)}]{ShayDrake:1998}%
  \BibitemOpen
  \bibfield  {author} {\bibinfo {author} {\bibnamefont {{Shay}}, \bibfnamefont
  {M.~A.}}\ and\ \bibinfo {author} {\bibnamefont {{Drake}}, \bibfnamefont
  {J.~F.}},\ }\bibfield  {title} {\enquote {\bibinfo {title} {{The role of
  electron dissipation on the rate of collisionless magnetic reconnection}},}\
  }\href {\doibase 10.1029/1998GL900036} {\bibfield  {journal} {\bibinfo
  {journal} {Geophysical Research Letters}\ }\textbf {\bibinfo {volume} {25}},\
  \bibinfo {pages} {3759--3762} (\bibinfo {year} {1998})}\BibitemShut {NoStop}%
\bibitem [{\citenamefont {{Shay}}\ \emph {et~al.}(1998)\citenamefont {{Shay}},
  \citenamefont {{Drake}}, \citenamefont {{Denton}},\ and\ \citenamefont
  {{Biskamp}}}]{Shay_etal:1998}%
  \BibitemOpen
  \bibfield  {author} {\bibinfo {author} {\bibnamefont {{Shay}}, \bibfnamefont
  {M.~A.}}, \bibinfo {author} {\bibnamefont {{Drake}}, \bibfnamefont {J.~F.}},
  \bibinfo {author} {\bibnamefont {{Denton}}, \bibfnamefont {R.~E.}}, \ and\
  \bibinfo {author} {\bibnamefont {{Biskamp}}, \bibfnamefont {D.}},\ }\bibfield
   {title} {\enquote {\bibinfo {title} {{Structure of the dissipation region
  during collisionless magnetic reconnection}},}\ }\href {\doibase
  10.1029/97JA03528} {\bibfield  {journal} {\bibinfo  {journal} {Journal of
  Geophysical Research}\ }\textbf {\bibinfo {volume} {103}},\ \bibinfo {pages}
  {9165--9176} (\bibinfo {year} {1998})}\BibitemShut {NoStop}%
\bibitem [{\citenamefont {{Sur}}\ and\ \citenamefont
  {{Brandenburg}}(2009)}]{SurBrandenburg:2009}%
  \BibitemOpen
  \bibfield  {author} {\bibinfo {author} {\bibnamefont {{Sur}}, \bibfnamefont
  {S.}}\ and\ \bibinfo {author} {\bibnamefont {{Brandenburg}}, \bibfnamefont
  {A.}},\ }\bibfield  {title} {\enquote {\bibinfo {title} {{The role of the
  Yoshizawa effect in the Archontis dynamo}},}\ }\href {\doibase
  10.1111/j.1365-2966.2009.15254.x} {\bibfield  {journal} {\bibinfo  {journal}
  {Monthly Notices of the Royal Astronomical Society}\ }\textbf {\bibinfo
  {volume} {399}},\ \bibinfo {pages} {273--280} (\bibinfo {year} {2009})},\
  \Eprint {http://arxiv.org/abs/0902.2394} {arXiv:0902.2394 [astro-ph.SR]}
  \BibitemShut {NoStop}%
\bibitem [{\citenamefont {{Sweet}}(1958)}]{Sweet:1958}%
  \BibitemOpen
  \bibfield  {author} {\bibinfo {author} {\bibnamefont {{Sweet}}, \bibfnamefont
  {P.~A.}},\ }\bibfield  {title} {\enquote {\bibinfo {title} {{The topology of
  force-free magnetic fields}},}\ }\href@noop {} {\bibfield  {journal}
  {\bibinfo  {journal} {The Observatory}\ }\textbf {\bibinfo {volume} {78}},\
  \bibinfo {pages} {30--32} (\bibinfo {year} {1958})}\BibitemShut {NoStop}%
\bibitem [{\citenamefont {{Takamoto}}, \citenamefont {{Inoue}},\ and\
  \citenamefont {{Lazarian}}(2015)}]{Takamoto_etal:2015}%
  \BibitemOpen
  \bibfield  {author} {\bibinfo {author} {\bibnamefont {{Takamoto}},
  \bibfnamefont {M.}}, \bibinfo {author} {\bibnamefont {{Inoue}}, \bibfnamefont
  {T.}}, \ and\ \bibinfo {author} {\bibnamefont {{Lazarian}}, \bibfnamefont
  {A.}},\ }\bibfield  {title} {\enquote {\bibinfo {title} {{Turbulent
  Reconnection in Relativistic Plasmas and Effects of Compressibility}},}\
  }\href {\doibase 10.1088/0004-637X/815/1/16} {\bibfield  {journal} {\bibinfo
  {journal} {The Astrophysical Journal}\ }\textbf {\bibinfo {volume} {815}},\
  \bibinfo {eid} {16} (\bibinfo {year} {2015})},\ \Eprint
  {http://arxiv.org/abs/1509.07703} {arXiv:1509.07703 [astro-ph.HE]}
  \BibitemShut {NoStop}%
\bibitem [{\citenamefont {{Webb}}\ \emph
  {et~al.}(2014{\natexlab{a}})\citenamefont {{Webb}}, \citenamefont
  {{Dasgupta}}, \citenamefont {{McKenzie}}, \citenamefont {{Hu}},\ and\
  \citenamefont {{Zank}}}]{Webb_etal:2014a}%
  \BibitemOpen
  \bibfield  {author} {\bibinfo {author} {\bibnamefont {{Webb}}, \bibfnamefont
  {G.~M.}}, \bibinfo {author} {\bibnamefont {{Dasgupta}}, \bibfnamefont {B.}},
  \bibinfo {author} {\bibnamefont {{McKenzie}}, \bibfnamefont {J.~F.}},
  \bibinfo {author} {\bibnamefont {{Hu}}, \bibfnamefont {Q.}}, \ and\ \bibinfo
  {author} {\bibnamefont {{Zank}}, \bibfnamefont {G.~P.}},\ }\bibfield  {title}
  {\enquote {\bibinfo {title} {{Local and nonlocal advected invariants and
  helicities in magnetohydrodynamics and gas dynamics I: Lie dragging
  approach}},}\ }\href {\doibase 10.1088/1751-8113/47/9/095501} {\bibfield
  {journal} {\bibinfo  {journal} {Journal of Physics A Mathematical General}\
  }\textbf {\bibinfo {volume} {47}},\ \bibinfo {eid} {095501} (\bibinfo {year}
  {2014}{\natexlab{a}})},\ \Eprint {http://arxiv.org/abs/1307.1105}
  {arXiv:1307.1105 [math-ph]} \BibitemShut {NoStop}%
\bibitem [{\citenamefont {{Webb}}\ \emph
  {et~al.}(2014{\natexlab{b}})\citenamefont {{Webb}}, \citenamefont
  {{Dasgupta}}, \citenamefont {{McKenzie}}, \citenamefont {{Hu}},\ and\
  \citenamefont {{Zank}}}]{Webb_etal:2014b}%
  \BibitemOpen
  \bibfield  {author} {\bibinfo {author} {\bibnamefont {{Webb}}, \bibfnamefont
  {G.~M.}}, \bibinfo {author} {\bibnamefont {{Dasgupta}}, \bibfnamefont {B.}},
  \bibinfo {author} {\bibnamefont {{McKenzie}}, \bibfnamefont {J.~F.}},
  \bibinfo {author} {\bibnamefont {{Hu}}, \bibfnamefont {Q.}}, \ and\ \bibinfo
  {author} {\bibnamefont {{Zank}}, \bibfnamefont {G.~P.}},\ }\bibfield  {title}
  {\enquote {\bibinfo {title} {{Local and nonlocal advected invariants and
  helicities in magnetohydrodynamics and gas dynamics: II. Noether's theorems
  and Casimirs}},}\ }\href {\doibase 10.1088/1751-8113/47/9/095502} {\bibfield
  {journal} {\bibinfo  {journal} {Journal of Physics A Mathematical General}\
  }\textbf {\bibinfo {volume} {47}},\ \bibinfo {eid} {095502} (\bibinfo {year}
  {2014}{\natexlab{b}})},\ \Eprint {http://arxiv.org/abs/1307.1038}
  {arXiv:1307.1038 [math-ph]} \BibitemShut {NoStop}%
\bibitem [{\citenamefont {{Widmer}}, \citenamefont {{B{\"u}chner}},\ and\
  \citenamefont {{Yokoi}}(2016{\natexlab{a}})}]{Widmer_etal:2016b}%
  \BibitemOpen
  \bibfield  {author} {\bibinfo {author} {\bibnamefont {{Widmer}},
  \bibfnamefont {F.}}, \bibinfo {author} {\bibnamefont {{B{\"u}chner}},
  \bibfnamefont {J.}}, \ and\ \bibinfo {author} {\bibnamefont {{Yokoi}},
  \bibfnamefont {N.}},\ }\bibfield  {title} {\enquote {\bibinfo {title}
  {{Characterizing plasmoid reconnection by turbulence dynamics}},}\ }\href
  {\doibase 10.1063/1.4962694} {\bibfield  {journal} {\bibinfo  {journal}
  {Physics of Plasmas}\ }\textbf {\bibinfo {volume} {23}},\ \bibinfo {eid}
  {092304} (\bibinfo {year} {2016}{\natexlab{a}})}\BibitemShut {NoStop}%
\bibitem [{\citenamefont {{Widmer}}, \citenamefont {{B{\"u}chner}},\ and\
  \citenamefont {{Yokoi}}(2016{\natexlab{b}})}]{Widmer_etal:2016a}%
  \BibitemOpen
  \bibfield  {author} {\bibinfo {author} {\bibnamefont {{Widmer}},
  \bibfnamefont {F.}}, \bibinfo {author} {\bibnamefont {{B{\"u}chner}},
  \bibfnamefont {J.}}, \ and\ \bibinfo {author} {\bibnamefont {{Yokoi}},
  \bibfnamefont {N.}},\ }\bibfield  {title} {\enquote {\bibinfo {title}
  {{Sub-grid-scale description of turbulent magnetic reconnection in
  magnetohydrodynamics}},}\ }\href {\doibase 10.1063/1.4947211} {\bibfield
  {journal} {\bibinfo  {journal} {Physics of Plasmas}\ }\textbf {\bibinfo
  {volume} {23}},\ \bibinfo {eid} {042311} (\bibinfo {year}
  {2016}{\natexlab{b}})},\ \Eprint {http://arxiv.org/abs/1511.04347}
  {arXiv:1511.04347 [physics.plasm-ph]} \BibitemShut {NoStop}%
\bibitem [{\citenamefont {{Widmer}}, \citenamefont {{B{\"u}chner}},\ and\
  \citenamefont {{Yokoi}}(2019)}]{Widmer_etal:2019}%
  \BibitemOpen
  \bibfield  {author} {\bibinfo {author} {\bibnamefont {{Widmer}},
  \bibfnamefont {F.}}, \bibinfo {author} {\bibnamefont {{B{\"u}chner}},
  \bibfnamefont {J.}}, \ and\ \bibinfo {author} {\bibnamefont {{Yokoi}},
  \bibfnamefont {N.}},\ }\bibfield  {title} {\enquote {\bibinfo {title}
  {{Analysis of fast turbulent reconnection with self-consistent determination
  of turbulence timescale}},}\ }\href {\doibase 10.1063/1.5109020} {\bibfield
  {journal} {\bibinfo  {journal} {Physics of Plasmas}\ }\textbf {\bibinfo
  {volume} {26}},\ \bibinfo {eid} {102112} (\bibinfo {year} {2019})},\ \Eprint
  {http://arxiv.org/abs/1905.01527} {arXiv:1905.01527 [physics.plasm-ph]}
  \BibitemShut {NoStop}%
\bibitem [{\citenamefont {{Woltjer}}(1958)}]{Woltjer:1958}%
  \BibitemOpen
  \bibfield  {author} {\bibinfo {author} {\bibnamefont {{Woltjer}},
  \bibfnamefont {L.}},\ }\bibfield  {title} {\enquote {\bibinfo {title} {{On
  Hydromagnetic Equilibrium}},}\ }\href {\doibase 10.1073/pnas.44.9.833}
  {\bibfield  {journal} {\bibinfo  {journal} {Proceedings of the National
  Academy of Science}\ }\textbf {\bibinfo {volume} {44}},\ \bibinfo {pages}
  {833--841} (\bibinfo {year} {1958})}\BibitemShut {NoStop}%
\bibitem [{\citenamefont {{Yamada}}, \citenamefont {{Kulsrud}},\ and\
  \citenamefont {{Ji}}(2010)}]{Yamada_etal.:2010}%
  \BibitemOpen
  \bibfield  {author} {\bibinfo {author} {\bibnamefont {{Yamada}},
  \bibfnamefont {M.}}, \bibinfo {author} {\bibnamefont {{Kulsrud}},
  \bibfnamefont {R.}}, \ and\ \bibinfo {author} {\bibnamefont {{Ji}},
  \bibfnamefont {H.}},\ }\bibfield  {title} {\enquote {\bibinfo {title}
  {{Magnetic reconnection}},}\ }\href {\doibase 10.1103/RevModPhys.82.603}
  {\bibfield  {journal} {\bibinfo  {journal} {Reviews of Modern Physics}\
  }\textbf {\bibinfo {volume} {82}},\ \bibinfo {pages} {603--664} (\bibinfo
  {year} {2010})}\BibitemShut {NoStop}%
\bibitem [{\citenamefont {{Yokoi}}(2013)}]{Yokoi:2013}%
  \BibitemOpen
  \bibfield  {author} {\bibinfo {author} {\bibnamefont {{Yokoi}}, \bibfnamefont
  {N.}},\ }\bibfield  {title} {\enquote {\bibinfo {title} {{Cross helicity and
  related dynamo}},}\ }\href {\doibase 10.1080/03091929.2012.754022} {\bibfield
   {journal} {\bibinfo  {journal} {Geophysical and Astrophysical Fluid
  Dynamics}\ }\textbf {\bibinfo {volume} {107}},\ \bibinfo {pages} {114--184}
  (\bibinfo {year} {2013})},\ \Eprint {http://arxiv.org/abs/1306.6348}
  {arXiv:1306.6348 [astro-ph.SR]} \BibitemShut {NoStop}%
\bibitem [{\citenamefont {{Yokoi}}(2018{\natexlab{a}})}]{Yokoi:2018a}%
  \BibitemOpen
  \bibfield  {author} {\bibinfo {author} {\bibnamefont {{Yokoi}}, \bibfnamefont
  {N.}},\ }\bibfield  {title} {\enquote {\bibinfo {title} {{Electromotive force
  in strongly compressible magnetohydrodynamic turbulence}},}\ }\href {\doibase
  10.1017/S0022377818000727} {\bibfield  {journal} {\bibinfo  {journal}
  {Journal of Plasma Physics}\ }\textbf {\bibinfo {volume} {84}},\ \bibinfo
  {eid} {735840501} (\bibinfo {year} {2018}{\natexlab{a}})}\BibitemShut
  {NoStop}%
\bibitem [{\citenamefont {{Yokoi}}(2018{\natexlab{b}})}]{Yokoi:2018b}%
  \BibitemOpen
  \bibfield  {author} {\bibinfo {author} {\bibnamefont {{Yokoi}}, \bibfnamefont
  {N.}},\ }\bibfield  {title} {\enquote {\bibinfo {title} {{Mass and
  internal-energy transports in strongly compressible magnetohydrodynamic
  turbulence}},}\ }\href {\doibase 10.1017/S0022377818001228} {\bibfield
  {journal} {\bibinfo  {journal} {Journal of Plasma Physics}\ }\textbf
  {\bibinfo {volume} {84}},\ \bibinfo {eid} {775840603} (\bibinfo {year}
  {2018}{\natexlab{b}})}\BibitemShut {NoStop}%
\bibitem [{\citenamefont {{Yokoi}}, \citenamefont {{Higashimori}},\ and\
  \citenamefont {{Hoshino}}(2013)}]{Yokoi_etal:2013}%
  \BibitemOpen
  \bibfield  {author} {\bibinfo {author} {\bibnamefont {{Yokoi}}, \bibfnamefont
  {N.}}, \bibinfo {author} {\bibnamefont {{Higashimori}}, \bibfnamefont {K.}},
  \ and\ \bibinfo {author} {\bibnamefont {{Hoshino}}, \bibfnamefont {M.}},\
  }\bibfield  {title} {\enquote {\bibinfo {title} {{Transport enhancement and
  suppression in turbulent magnetic reconnection: A self-consistent turbulence
  model}},}\ }\href {\doibase 10.1063/1.4851976} {\bibfield  {journal}
  {\bibinfo  {journal} {Physics of Plasmas}\ }\textbf {\bibinfo {volume}
  {20}},\ \bibinfo {eid} {122310} (\bibinfo {year} {2013})},\ \Eprint
  {http://arxiv.org/abs/1401.1498} {arXiv:1401.1498 [physics.plasm-ph]}
  \BibitemShut {NoStop}%
\bibitem [{\citenamefont {{Yokoi}}\ and\ \citenamefont
  {{Hoshino}}(2011)}]{YokoiHoshino:2011}%
  \BibitemOpen
  \bibfield  {author} {\bibinfo {author} {\bibnamefont {{Yokoi}}, \bibfnamefont
  {N.}}\ and\ \bibinfo {author} {\bibnamefont {{Hoshino}}, \bibfnamefont
  {M.}},\ }\bibfield  {title} {\enquote {\bibinfo {title} {{Flow-turbulence
  interaction in magnetic reconnection}},}\ }\href {\doibase 10.1063/1.3641968}
  {\bibfield  {journal} {\bibinfo  {journal} {Physics of Plasmas}\ }\textbf
  {\bibinfo {volume} {18}},\ \bibinfo {pages} {111208--111208} (\bibinfo {year}
  {2011})},\ \Eprint {http://arxiv.org/abs/1105.6343} {arXiv:1105.6343
  [astro-ph.SR]} \BibitemShut {NoStop}%
\bibitem [{\citenamefont {{Yoshizawa}}(1990)}]{Yoshizawa:1990}%
  \BibitemOpen
  \bibfield  {author} {\bibinfo {author} {\bibnamefont {{Yoshizawa}},
  \bibfnamefont {A.}},\ }\bibfield  {title} {\enquote {\bibinfo {title}
  {{Self-consistent turbulent dynamo modeling of reversed field pinches and
  planetary magnetic fields}},}\ }\href {\doibase 10.1063/1.859484} {\bibfield
  {journal} {\bibinfo  {journal} {Physics of Fluids B}\ }\textbf {\bibinfo
  {volume} {2}},\ \bibinfo {pages} {1589--1600} (\bibinfo {year}
  {1990})}\BibitemShut {NoStop}%
\bibitem [{\citenamefont {{Zweibel}}\ and\ \citenamefont
  {{Yamada}}(2009)}]{ZweibelYamada:2009}%
  \BibitemOpen
  \bibfield  {author} {\bibinfo {author} {\bibnamefont {{Zweibel}},
  \bibfnamefont {E.~G.}}\ and\ \bibinfo {author} {\bibnamefont {{Yamada}},
  \bibfnamefont {M.}},\ }\bibfield  {title} {\enquote {\bibinfo {title}
  {{Magnetic Reconnection in Astrophysical and Laboratory Plasmas}},}\ }\href
  {\doibase 10.1146/annurev-astro-082708-101726} {\bibfield  {journal}
  {\bibinfo  {journal} {Annual Review of Astronomy And astrophysics}\ }\textbf
  {\bibinfo {volume} {47}},\ \bibinfo {pages} {291--332} (\bibinfo {year}
  {2009})}\BibitemShut {NoStop}%
\end{thebibliography}%

\end{document}